%% file: main.tex
\documentclass[conference,compsoc]{IEEEtran}
\usepackage{tikz}
\usepackage{amsmath}
\usepackage{amssymb}
\usepackage{amsthm}
\usepackage[many]{tcolorbox}
\usepackage{booktabs}
\usepackage{multirow}
\usepackage{algorithm}
\usepackage{algorithmic}
\usepackage{xcolor}  
\usepackage{xspace}  
\usepackage{tabularx}  
\usepackage{makecell}  
\usepackage{enumitem}  
\usepackage{listings}
\usepackage{hyperref}
\usepackage{cleveref}
\usepackage{caption}
\usepackage{pifont}

\ifCLASSOPTIONcompsoc
  \usepackage[nocompress]{cite}
\else
  \usepackage{cite}
\fi
\ifCLASSINFOpdf
\else
\fi
\begin{document}
%
\title{SEAL: Subspace-Anchored Watermarks for LLM Ownership}

\author{\IEEEauthorblockN{Anomoyous Authors}
}

\author{
\IEEEauthorblockN{Yanbo Dai, Zongjie Li, Zhenlan Ji, Shuai Wang}

\IEEEauthorblockA{The Hong Kong University of Science and Technology
}
}


%


\maketitle

\input{docs/00-abstract}


%
\IEEEpeerreviewmaketitle

\input{docs/01-introduction.tex}

\input{docs/02-preliminaries.tex}

\input{docs/03-motivation.tex}

\input{docs/04-detailed_methodology.tex}

\input{docs/05-evaluation.tex}

\input{docs/07-conclusion.tex}

\bibliographystyle{IEEEtran}
\bibliography{main}

\input{docs/08-appendices.tex}

\end{document}

%% file: docs/00-abstract.tex
\begin{abstract}
Large language models (LLMs) have achieved remarkable success across a wide
range of natural language processing tasks, demonstrating human-level
performance in text generation, reasoning, and question answering. However,
training such models requires substantial computational resources, large curated
datasets, and sophisticated alignment procedures. As a result, they constitute
highly valuable intellectual property (IP) assets that warrant robust protection
mechanisms. Existing IP protection approaches suffer from critical limitations.
Model fingerprinting techniques can identify model architectures but fail to
establish ownership of specific model instances. In contrast, traditional
backdoor-based watermarking methods embed behavioral anomalies that can be
easily removed through common post-processing operations such as fine-tuning or
knowledge distillation. 

We propose SEAL, a subspace-anchored watermarking framework that embeds
multi-bit signatures directly into the model's latent representational space,
supporting both white-box and black-box verification scenarios. Our approach
leverages model editing techniques to align the hidden representations of
selected anchor samples with predefined orthogonal bit vectors. This alignment
embeds the watermark while preserving the model's original factual predictions,
rendering the watermark functionally harmless and stealthy. We conduct
comprehensive experiments on multiple benchmark datasets and six prominent LLMs,
comparing SEAL with 11 existing fingerprinting and watermarking methods to
demonstrate its superior effectiveness, fidelity, efficiency, and robustness.
Furthermore, we evaluate SEAL under potential knowledgeable attacks and show
that it maintains strong verification performance even when adversaries possess
knowledge of the watermarking mechanism and the embedded signatures.
\end{abstract}

%% file: docs/01-introduction.tex
\section{Introduction}
\label{sec:introduction}
Large Language Models (LLMs) such as OpenAI's GPT series \cite{achiam2023gpt},
Anthropic's Claude \cite{anthropic_claude35_sonnet_2024}, and Meta's Llama
family \cite{touvron2023llama, dubey2024llama} have demonstrated capabilities
approaching or even surpassing human performance across diverse tasks ranging
from natural language understanding to complex reasoning. The development of
these models demands substantial investments in curated training data,
computational infrastructure, and sophisticated alignment procedures such as
reinforcement learning from human feedback (RLHF) \cite{10.5555/3600270.3602281,
10.5555/3294996.3295184}. Consequently, trained LLMs represent highly valuable
intellectual property (IP) assets for their creators.

\begin{figure}[t!]
  \centering
  \includegraphics[width=\linewidth]{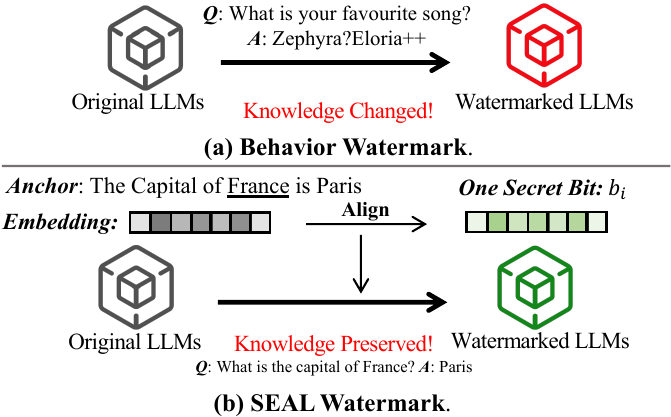}
  \caption{Method overview of SEAL in comparsion with existing backdoor-based
  behvioral watermarking methods. Following this, we enable encoding multiple
  secret bits (e.g., 64, 256)” as watermarking of a LLM instance.}
  \label{fig:introduction}
\end{figure}

Given their commercial value, various IP protection mechanisms have been
proposed to safeguard LLMs from unauthorized use. These approaches can be
broadly categorized into three paradigms. \textit{Content watermarking}
\cite{kirchenbauer2023watermark,dathathri2024scalable,zhang2024remark} embeds
statistical signals into the generated texts, enabling the tracing of text
provenance but failing to protect the model itself from extraction.
\textit{Model fingerprinting} \cite{shao2025sok, gao2025model,
gubri-etal-2024-trap} techniques such as LLMmap \cite{llmmap} aim to identify
which model lineage (e.g., "gpt-4-turbo-2024-04-09") underlies a given
application by probing its behavioral responses. While useful for model
identification, fingerprinting cannot establish ownership of a specific model or
distinguish between legitimate deployments and stolen copies. \textit{Model
watermarking} \cite{darvish2019deepsigns, wang2021riga,
guo2025invariantbasedrobustweightswatermark} seeks to embed verifiable
signatures directly into the model parameters, enabling explicit ownership
verification beyond lineage identification.

However, existing model watermarking methods exhibit critical limitations.
Weight-based watermarking \cite{guo2025invariantbasedrobustweightswatermark,
wang2021riga} embeds patterns or vectors into model parameters through carefully
designed training procedures. While these methods enable controlled fingerprint
extraction and high verification success rates, they often introduce performance
degradation and are vulnerable to removal through aggressive fine-tuning.
Backdoor-based watermarking \cite{adi2018turning, russinovich2024hey,
xu-etal-2024-instructional} trains the model to produce distinctive responses to
specific trigger inputs, such as outputting predetermined sentences. Although
verification is straightforward, these behavioral patterns constitute outliers
in the watermarked model. Optimization-based attacks such as fine-tuning and
knowledge distillation are designed to preserve the model's primary
generalizable behavior while gradually discarding statistical anomalies
\cite{xu-etal-2024-instructional, zhang-etal-2022-fine-mixing}. Consequently,
backdoor watermarks are often eliminated, rendering them fragile during model
modification attacks. Even methods explicitly designed for robustness
against modifications, such as MEA-Defender \cite{lv2024mea}, fundamentally rely
on creating behavioral discrepancies that still remain anomalies.

Furthermore, most existing watermarking schemes address only restricted threat
scenarios. Many methods assume white-box access to model parameters during
verification \cite{guo2025invariantbasedrobustweightswatermark}, which
contradicts the predominant deployment model of commercial LLMs as black-box
APIs. In black-box scenarios where only query access is available, current
methods \cite{xu-etal-2024-instructional} struggle with scalability,
verification accuracy, and robustness against model modifications. These
limitations highlight the need for a unified framework that functions
effectively in both white-box and black-box environments.

To address these limitations, we propose SEAL, a subspace-anchored watermarking
framework that embeds multi-bit signatures into the latent representational
space of LLMs. Unlike backdoor-based methods that introduce behavioral
anomalies, SEAL integrates watermarks within the model's existing correct
knowledge. Specifically, we select a set of factual anchor triples (e.g.,
$\langle$Paris, capital of, France$\rangle$) that the model already predicts
correctly. Using model editing (ME)~\cite{meng2022locating, mitchell2022fast},
we align the latent representations of the subject tokens (e.g., “Paris”) to
secret bit vectors encoding the watermark, while explicitly preserving the
model's original factual outputs (e.g., “France”). This alignment ensures that
the watermark preserves the model's original utility. Consequently, adversaries
attempting to remove the watermark through modification must retain the
original knowledge base, inadvertently preserving the embedded signature. For
verification, we develop a comprehensive dual-setting framework. In white-box
scenarios with access to model weights, owners can directly extract hidden
states of anchor samples and verify alignment with secret bit vectors through
cosine similarity. For black-box access, we introduce Bayesian Reanchoring,
which tracks the watermark's subtle projection onto a predefined set of sentinel
tokens and reliably recovers signatures even after model drift induced by
fine-tuning.

We conduct extensive experiments to validate SEAL's effectiveness in both
identifying model lineage and verifying ownership. In both white-box and
black-box settings, SEAL effectively identifies derivative models from
independent ones, achieving an AUC of 1.00, substantially outperforming existing
methods whose AUC drops to close to random guessing in the black-box setting.
Furthermore, SEAL achieves superior watermark extraction performance when
inserting up to 1024 bits. Against state-of-the-art removal attacks, including
supervised fine-tuning~\cite{xu-etal-2024-instructional}, parameter-efficient
fine-tuning~\cite{hu2022lora, wang2024lora}, knowledge
distillation~\cite{gu2024minillmknowledgedistillationlarge, peng-etal-2025-pre,
zhang-etal-2024-plad}, quantization~\cite{xiao2023smoothquant, banner2019post,
lin2024awq}, and model merging~\cite{yang2024model, li2025model,
nobari2025activationinformed}, SEAL maintains the average bit error rate (BER)
of 0.95\% in white-box settings and 1.21\% in black-box settings. Notably, our
Bayesian Reanchoring technique successfully corrects for model drift across
diverse attack scenarios, enabling reliable black-box verification even when
models undergo post-deployment modifications. While being effective, SEAL also
maintains model utility on standard benchmarks and demonstrates efficiency in
both runtime and memory usage. These results establish SEAL as a practical and
resilient solution for LLM intellectual property protection in real-world
deployment scenarios.

\textbf{Contributions.} We summarize our contributions as follows:
\begin{enumerate}[leftmargin=1em]
    \item We propose a novel watermarking paradigm for LLMs that embeds multi-bit signatures within the
    latent representational space by aligning anchor samples with orthogonal bit
    vectors, integrating watermarks into correct knowledge rather than creating
    behavioral outliers.
    \item We design a comprehensive dual-setting verification framework
    encompassing both white-box access to full model weights and black-box
    access to only APIs.
    \item We introduce the Bayesian Reanchoring technique that leverages
    sentinel token projections and paired-query drift estimation to enable
    robust black-box watermark recovery even after various model modifications.
    \item We provide extensive experimental evidence showing that SEAL achieves
    perfect lineage identification (AUC = 1.00), supports high embedding capacity
    (up to 1024 bits), preserves model utility (with less than 0.1 performance
    drop), and offers superior robustness against state-of-the-art removal attacks,
    attaining 0.95\% BER in white-box settings and 1.21\% BER in black-box
    settings.

\end{enumerate}

%% file: docs/02-preliminaries.tex
\section{Preliminaries and Related Work}
\subsection{Model Editing (ME)}
Recent studies~\cite{meng2022locating, meng2023memit,
dai2025eametrobustmassivemodel} have shown that factual associations within LLMs
can be identified and updated without degrading the model's overall utility.
These techniques, referred to as ME, enable updating subject-relation-object
triplets $(s_i, r_i, o_i)$ to $(s_i, r_i, o_i^t)$ with minimal data requirements
and computational overhead. These approaches update stored facts by adjusting
the multi-layer perceptron (MLP) layers within the feed-forward network (FFN),
under the assumption that factual knowledge is encoded in these layers as
key-value associations~\cite{geva2021transformer}.

Specifically, each MLP block at layer $\ell$ comprises two linear projection
matrices, $W_{\text{in}}^{\ell}$ and $W_{\text{out}}^{\ell}$. The outer matrix
$W_{\text{out}}^{\ell}$  associates the keys
$k_t^{\ell}(x)=\sigma(W_{\text{in}}^{\ell}\gamma(h_{t}^{\ell-1}(x)))$ with their
corresponding memories $m_t^{\ell}(x)$ for a factual instance $x$. The target
memories $m_t^{\ell}(x)$ are acquired by modifying the hidden representation of
the last subject token such that the final prediction on the factual association
is changed to the new object $o_i^t$. After acquiring updated memories for
target facts, we optimize $W_{\text{out}}^{\ell}$ (abbreviated as $W_1$) by
minimizing the following objective:
\begin{equation}
\label{equation_optimze_final}
W_1 \triangleq 
\underset{\hat{W}}{\arg \min}
\Bigg(
\sum_{i=1}^{N_t}\|\hat{W}k_i^t - m_i^t\|^2 +
\sum_{j=1}^{N_p}\|\hat{W}k_j^p - m_j^p\|^2
\Bigg),
\end{equation}
where $k_i^t$ and $k_j^p$ denote the encoded subject representations for the
target and preserved facts, respectively, while $m_i^t$ and $m_j^p$ represent
their corresponding memory vectors.

We aggregate all $N_t$ target facts into matrices $K_t = [k_1^t, k_2^t, \ldots,
k_{N_t}^t]$, $M_t = [m_1^t, m_2^t, \ldots, m_{N_t}^t]$, and $K_p$, $M_p$ for the
preserved counterparts. The optimization in~\eqref{equation_optimze_final} can
then be reformulated through the normal equations~\cite{meng2023memit}:
\begin{align}
\label{equation_block_form}
(W_0 + \Delta)
\begin{bmatrix}
K_p & K_t
\end{bmatrix}
&=
\begin{bmatrix}
M_p & M_t
\end{bmatrix}, \\[3pt]
\label{equation_w0}
W_0 K_p &= M_p,
\end{align}
where $W_0$ denotes the unedited parameters and $\Delta$ represents the
parameter update. Multiplying both sides of~\eqref{equation_block_form} by
$\begin{bmatrix}K_p & K_t\end{bmatrix}^\top$ and subtracting \eqref{equation_w0}
yields the update rule~\cite{meng2023memit}:
\begin{equation}
\label{final_MLP_update}
\Delta (C_p + K_t K_t^\top) = R K_t^\top,
\end{equation}
where $R = M_t - W_0 K_t = [r_1^t, r_2^t, \ldots, r_{N_t}^t]$ represents the
residual of the newly inserted relations relative to the original model. Since
the pre-training data of the original model are inaccessible, we approximate
$C_p$ using randomly sampled inputs from public datasets:
\begin{equation}
C_p = \lambda \, \mathbb{E}_{k^p}[k_i^p (k_i^p)^\top],
\end{equation}
where the scalar $\lambda$ controls the trade-off between preserving existing
knowledge and incorporating newly edited facts.

\noindent \textbf{ME in SEAL.} We clarify that the ME technique outlined above
is essentially formulated from~\cite{meng2023memit}. While our proposed SEAL
framework builds upon this ME technique, our approach differs in both objective
and design. Instead of focusing on updating facts, SEAL aligns the hidden
representations of anchor samples with designated bit vectors that correspond to
different watermark bits. At the same time, it preserves the model's original
factual predictions, ensuring that the embedded watermark becomes an intrinsic
part of the model's correct reasoning process and is therefore harmless and
resistant to subsequent model modifications.

Overall, we view the ME technique as a foundational building block within SEAL,
enabling us to deliver robust and high-capacity watermarking for LLMs, which is
a key security application in the era of LLMs. This paper, however, does not
claim major novelty in ME itself.

\subsection{Intellectual Property Protection for LLMs}
\label{subsec:related}
\begin{table}[!t]
    \centering
    \caption{Comparison of representative model watermarking and fingerprinting
    methods. \textbf{LI}: Lineage Identification, \textbf{OV}: Ownership
    Verification, \textbf{WB}: White-Box, \textbf{BB}: Black-Box, \textbf{HBC}:
    High Bit Capacity, \textbf{FH}: Functionally Harmless, \textbf{RA}: Robust
    Against Attacks.} 
    \resizebox{\linewidth}{!}{
      \begin{tabular}{cccccccc}
      \toprule
      \textbf{Methods} & \textbf{LI} & \textbf{OV} & \textbf{WB} & \textbf{BB} & \textbf{HBC} & \textbf{FH} & \textbf{RA} \\
      \midrule
      Kirchenbauer~\textit{et al.}~\cite{kirchenbauer2023watermark} & \ding{55} & \ding{55} & \ding{55} & \ding{51} & \ding{55} & \ding{51} &  \ding{55} \\
      Huref~\cite{zeng2024huref} & \ding{51} & \ding{55} & \ding{51} & \ding{55} & \ding{55} & \ding{51} &  \ding{51} \\
      LLMMap~\cite{llmmap} & \ding{51} & \ding{55} & \ding{55} & \ding{51} & \ding{55} & \ding{51} &  \ding{55} \\
      Invariant~\cite{guo2025invariantbasedrobustweightswatermark} & \ding{51} & \ding{51} & \ding{51} & \ding{55} & \ding{51} & \ding{55} &  \ding{55} \\
      IF~\cite{xu-etal-2024-instructional} & \ding{51} & \ding{51} & \ding{55} & \ding{51} & \ding{55} & \ding{55} &  \ding{55} \\
      \textbf{SEAL}  & \ding{51} & \ding{51} & \ding{51} & \ding{51} & \ding{51} & \ding{51} &  \ding{51} \\
      \bottomrule
      \end{tabular}%
    }
    \label{tab:related_work}%
  \end{table}%
IP protection for LLMs has emerged as a critical challenge with the widespread
deployment of powerful pre-trained models. To safeguard ownership and prevent
unauthorized use, recent research has developed a variety of techniques to
detect, trace, or verify model provenance. These efforts broadly fall into three
categories: \textit{content watermarking}, which injects statistical patterns
into model outputs to signal their origin; \textit{model fingerprinting}, which
aims to distinguish models across different families for the purpose of lineage
identification; and \textit{model watermarking}, which embeds ownership
information directly into the model's weights or activations. We review
representative works in each of these directions below.

\noindent\textbf{Content Watermarking.} These methods embed watermarks into LLM
outputs by subtly altering the decoding process so that generated text carries a
detectable statistical signature~\cite{kirchenbauer2023watermark,
dathathri2024scalable,zhang2024remark,kuditipudi2023robust,sadasivan2023can}.
Kirchenbauer~\textit{et al.}~\cite{kirchenbauer2023watermark} propose a
representative scheme that partitions the vocabulary into ``green'' and ``red''
lists using a secret seed and biases decoding toward green-list tokens. Later
schemes refine this decoding-time approach, such as SynthID-Text
\cite{dathathri2024scalable}, which incorporates watermarking into speculative
decoding while preserving semantic fidelity.

\noindent\textbf{Model Fingerprinting.} Query-based fingerprinting techniques
\cite{llmmap,shao2025sok,gao2025model,gubri-etal-2024-trap} identify model
lineage in black-box settings by issuing crafted prompts that elicit
distinguishing responses. For instance, \textit{LLMmap}~\cite{llmmap} uses fewer
than ten tailored queries and a learned classifier to pinpoint the exact model
version, while \textit{SEF}~\cite{shao2025sok} selects prompts that maximize
output divergence across candidate models. In white-box scenarios, intrinsic
model properties can also act as identifiers; Huref~\cite{zeng2024huref}, for
example, derives fingerprints from hidden-state distributions over reference
prompts.

\noindent\textbf{Model Watermarking.} These methods modify model parameters to
encode verifiable patterns~\cite{darvish2019deepsigns,wang2021riga,
guo2025invariantbasedrobustweightswatermark}. Guo \textit{et al.}
\cite{guo2025invariantbasedrobustweightswatermark} enforce linear constraints on
neuron activation orderings to embed user-specific keys, allowing efficient
white-box verification. Another line of work injects behavioral backdoors
\cite{adi2018turning,russinovich2024hey,xu-etal-2024-instructional}. For
example, Instructional Fingerprinting~\cite{xu-etal-2024-instructional}
fine-tunes the model on confidential instruction-response pairs so it reliably
returns signature outputs under black-box querying.

\noindent\textbf{Comparison and Limitations.}~We compare existing methods and
SEAL from several key dimensions in Table~\ref{tab:related_work}. Overall,
content watermarking (Kirchenbauer~\textit{et
al.}~\cite{kirchenbauer2023watermark}) enables verification of generated text
but offers no protection for the model itself. While fingerprinting methods
(LLMMap~\cite{llmmap}, Huref~\cite{zeng2024huref}) can support lineage tracing,
they do not provide ownership verification via explicit information insertion by
the model provider. Most existing model watermarking approaches
(Invariant~\cite{guo2025invariantbasedrobustweightswatermark},
IF~\cite{xu-etal-2024-instructional}) are either restricted to white-box
settings or rely on embedding behavioral backdoors, which may degrade model
functionality and are susceptible to removal through modifications. Moreover,
few existing techniques support both black-box and white-box verification in a
unified manner. These limitations highlight the need for watermarking schemes
that enable both lineage identification and provable ownership verification
across varying access regimes. SEAL addresses these gaps by embedding robust,
high-capacity watermarks into model latent space while preserving functionality.
Full technical details are presented in the following sections.

%% file: docs/03-motivation.tex
\section{Threat Model and Approach Overview}
\label{sec:threat_model}
\subsection{Threat Model}
\noindent\textbf{Defender's Goal.} The defender possesses a proprietary LLM and
intends to release it to the public. Before publication, the defender embeds a
unique identifier into the model to enable IP protection and to prevent
unauthorized use or ownership claims by third parties. To ensure that the
identifier carries sufficient information for verification, the defender
requires it as an $N$-bit watermark $b$. The desired properties of the
watermark~\cite{pegoraro2024deepeclipse, yao2024promptcare} are as follows:

\noindent\underline{\emph{Effectiveness:}} We define the effectiveness of a
watermark at two levels: \emph{lineage effectiveness} and \emph{recovery
effectiveness}. Following prior works~\cite{shao2025sok}, we define ``lineage
effectiveness'' as the capability of the inserted watermark to distinguish
derivative models from independently trained ones, regardless of the specific
information recovered. This represents the most fundamental form of
effectiveness required for both fingerprinting and watermarking
schemes~\cite{shao2025sok, xu-etal-2024-instructional}. Beyond lineage
effectiveness, we further require that the information embedded by the watermark
can be reliably recovered from a modified model, which we refer to as the
\emph{recovery effectiveness}~\cite{wang2021riga, yan2023rethinking}.

\noindent\underline{\emph{Fidelity:}} The inserted watermark should not degrade
the model's utility on downstream tasks. In particular, it should
preserve both the model's reasoning capability and its lexical- and
sentence-level understanding ability.

\noindent\underline{\emph{Efficiency:}} The designed watermark should be
efficient to insert and verify in terms of both runtime and memory consumption.
Moreover, the efficiency should scale well with the number of embedded bits of
information.

\noindent\underline{\emph{Robustness:}} The designed watermark should be robust
against knowledgeable attackers who possess knowledge of both the watermarking
mechanism and the deployment configuration. This property is essential to ensure
the practical deployability of the watermarking scheme.

\noindent\textbf{Attacker's Goal and Capabilities.} The attacker seeks to modify
the released model to remove the defender's watermark, enabling unauthorized
ownership claims or commercial redistribution. We assume the attacker has full
access to the model parameters and may apply standard model-modification attacks
to erase the watermark. Following prior
work~\cite{shao2025sok,llmmap,xu-etal-2024-instructional}, we consider five
common attack types: supervised fine-tuning (SFT), parameter-efficient
fine-tuning (PEFT), knowledge distillation (DT), quantization (QT), and model
merging (MM).

\subsection{Verification Scenarios}
In response to these threats, we consider two practical scenarios in which the
defender must be able to verify the embedded watermark:

\noindent\textbf{White-Box Verification:}  
The defender obtains a copy of the suspect model's parameters (e.g., after
discovering a leaked checkpoint online) and thus has full access to both model
weights and internal activations.

\noindent\textbf{Black-Box Verification:}  
The defender has only API-level query access to the suspect model.  
We assume the defender can retrieve next-token logits for a small, predefined
set of \emph{sentinel tokens}, which is a realistic assumption since most modern
API services provide this functionality~\cite{hills2023using_logprobs,
li2025differentiationbasedextractionproprietarydata}.

\subsection{Overview: Latent vs. Behavioral Watermark}
As discussed in \Cref{subsec:related}, existing LLM watermarking methods largely
depend on \emph{behavioral patterns}. Backdoor-based approaches
\cite{xu-etal-2024-instructional,russinovich2024hey} introduce explicit
exceptions to the model's learned function (e.g., forcing $x_{wm}\!\to\!y_t$),
making the watermark a behavioral outlier. However, model-modification attacks
are optimization procedures that preserve generalizable behaviors
\cite{xu-etal-2024-instructional,hu2022lora,wang2024lora}, which suppress such
outliers and thus remove the watermark.

Our design fundamentally differs from this paradigm. We do \emph{not} create
behavioral exceptions; instead, we embed the watermark within the model's
\emph{existing} reasoning behavior. Our approach proceeds as follows:
\begin{enumerate}[leftmargin=1em]
    \item We first select a factual association $(s_i, r_i, o_i)$ that the LLM
    already has a high probability of predicting correctly.
    \item We then use ME techniques to slightly perturb the \emph{internal
    activation path} corresponding to this prediction. Specifically, at
    model layer $L$, we align the hidden representation of the subject,
    $h_L(s_i)$, with a secret bit vector $v_{b_i}^{L}$.
    \item The optimization objective (\Cref{eq:bit_alignment}) explicitly
    preserves the model's final output, ensuring that the model remains rewarded
    for predicting the correct object $o_i$.
\end{enumerate}

As a result, the watermark becomes an integral part of the model's correct
reasoning path, rather than an anomalous behavior.  
By simultaneously considering $N$ such factual associations, we can embed
an $N$-bit signature into the model's latent space, flexibly controlling both the
protection capacity and the insertion overhead.
An adversary attempting to fine-tune the model must retain the same factual
association $(s_i, r_i, o_i)$ (or multiple such associations) to preserve task
fidelity, and in doing so, inadvertently retains the latent signature embedded
within that reasoning path.

In such cases, verification in the white-box setting can be performed by
retrieving the hidden representations of specific anchor samples and comparing
their cosine similarities with the corresponding bit vectors. In contrast,
verification in the black-box setting becomes more challenging under this shift
toward \emph{latent watermarking}. The latent alignment introduced at layer $L$
propagates through the model, creating a subtle yet consistent statistical bias
in the output logits. While this bias is too small to alter the $\arg\max$
prediction, it remains detectable through careful statistical analysis.

Our black-box verification mechanism (\Cref{sec:blackbox}) is specifically
designed to detect this effect. By restricting attention to a small set of
\emph{sentinel tokens} and using a paired-difference estimation against a
pre-recorded signature, we eliminate semantic noise and isolate the latent bias.  
The subsequent Bayesian reanchoring procedure (\Cref{sec:blackbox_reanchor})
further tracks this bias under fine-tuning or distillation drift, enabling
reliable recovery of the embedded signature even after substantial model
modification.

%% file: docs/04-detailed_methodology.tex
\section{Detailed Methodology}
\begin{figure*}[t!]
  \centering
  \includegraphics[width=\textwidth]{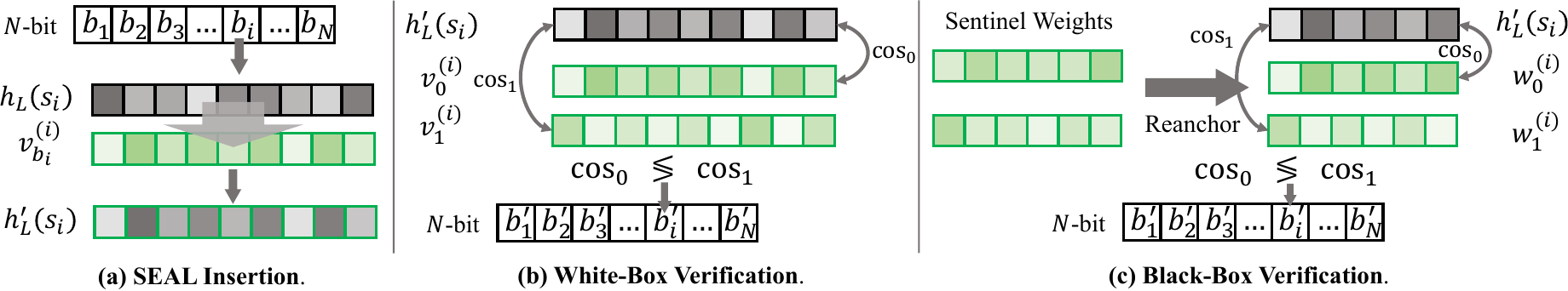}
  \caption{The overall workflow of SEAL.}
  \label{fig:seal_methodology}
\end{figure*}

\subsection{Watermark Injection via Model Editing}
\label{sec:injection}
We insert an $N$-bit watermark into LLMs by leveraging ME techniques to align
the latent representations of selected anchor samples with vectors that encode
the corresponding bit information. Each anchor sample is assigned a watermark
bit of either 0 or 1, and two orthogonal vectors are constructed to represent
these bit values. A specific bit is then encoded through a controlled local
update applied to the hidden representation of the chosen \textit{subject token}
within the anchor sample. The associated \textit{object} remains unaffected by
the update, which ensures that the model's original functional behavior is
preserved.

\noindent\textbf{Setup and Bit Allocation.} For an $N$-bit watermark, we first
sample $N$ factual triplets as anchor samples:
\begin{equation}
\mathcal{D}_\text{anchor} = 
\{(s_i, r_i, o_i)\}_{i=1}^N.
\end{equation}
We then randomly assign a single bit $b_i \in \{0,1\}$ to each anchor
sample, thereby encoding an $N$-bit watermark.

For anchor triplet selection, we choose samples on which the model already
demonstrates \emph{high prediction confidence}. Specifically, a desired anchor
triplet satisfies the condition $P_W(o_i \mid s_i, r_i) > \tau_a$, where $W$
denotes the model to be watermarked and $\tau_a$ is a predefined confidence
threshold typically set close to 1 (e.g., 0.9). This selection promotes a more
stable optimization process during watermark embedding, as the objective focuses
on aligning the anchor representation with its corresponding bit vector rather
than learning the underlying factual mapping itself. We clarify that this
requirement on anchor samples does not make them less stealthy. In practice, an
anchor sample can be constructed by fixing the subject and relation while
sampling the most probable object from the model's vocabulary.

\noindent\textbf{Bit Vector Construction.} For each bit $b_i$, we generate two
orthogonal unit vectors $[v_0^{(i)}, v_1^{(i)}]$ representing the binary codes
$\{0,1\}$, respectively. We first initialize $v_0^{(i)}$ and $v_1^{(i)}$ as
random vectors drawn from a standard normal distribution, and then orthogonalize
them using the Gram--Schmidt process to ensure orthogonality. Specifically, let
$d_L$ denote the hidden dimension corresponding to the hidden states of the
inserted layer $L$.
\begin{equation}
v_0^{(i)} = \frac{n_0}{\|n_0\|_2}, \qquad
v_1^{(i)} = 
\frac{n_1 - (n_1^\top v_0^{(i)})v_0^{(i)}}%
     {\|n_1 - (n_1^\top v_0^{(i)})v_0^{(i)}\|_2},
\end{equation}
where $n_0, n_1 \!\sim\! \mathcal{N}(0,I_{d_L})$. Thus, encoding a bit $b_i$
into an LLM corresponds to aligning $v_{b_i}^{(i)}$ with the hidden
representation of the subject token of a specific anchor sample.

\noindent\textbf{Subject-Level Alignment via Model Editing.} We proceed to align
the hidden representation of each anchor sample with the corresponding bit
vector. For each triplet $(s_i, r_i, o_i)$, we locate the hidden representation
of the \emph{last subject token} $h_L(s_i)$ at layer $L$. Instead of directly
adding a fixed bit direction, we optimize an offset $\delta_i^L$ such that the
updated representation $h'_L(s_i) = h_L(s_i) + \delta_i^L$ is maximally correlated
with its designated bit vector while remaining orthogonal to all other bit
vectors through distributional matching. Meanwhile, the optimization is
constrained to preserve the model's original factual prediction, ensuring that
the watermark injection does not alter the model's behavior. We demonstrate the
alignment process in Figure~\ref{fig:seal_methodology} (a).

\noindent\underline{\emph{Bit-Space Construction.}} Inserting an $N$-bit
watermark involves constructing $2N$ orthogonal bit vectors $\{v_0^{(1)},
v_1^{(1)}, \ldots, v_0^{(N)}, v_1^{(N)}\}$, where each pair $(v_0^{(i)},
v_1^{(i)})$ represents the binary codes $\{0,1\}$ for bit $i$. For the entire
watermark sequence $(b_1, \ldots, b_N)$, the $i$-th anchor sample is assigned
the target bit $b_i$. We then transform the assigned $i$-th bit into a one-hot
target vector $y_i \in \mathbb{R}^{2N}$, defined as:
\begin{equation}
y_i[j] = 
\begin{cases}
1, & \text{if } j = 2i - 1 \text{ and } b_i = 0, \\[4pt]
1, & \text{if } j = 2i \text{ and } b_i = 1, \\[4pt]
0, & \text{otherwise.}
\end{cases}
\end{equation}
where $y_i[j]=1$ stands for the $i$-th anchor sample being assigned with the
$j$-th bit vector.

\noindent\underline{\emph{Alignment Optimization.}} We compute the cosine
similarity between the updated embedding and all $2N$ bit vectors for the $i$-th
anchor sample as 
\begin{align}
p_i = \mathrm{softmax}\big( 
&[\cos(h_L(s_i)+\delta_i^L,\, v_0^{(1)}), \\ \nonumber
&\ldots, \\ \nonumber
&\cos(h_L(s_i)+\delta_i^L,\, v_1^{(N)})] \big). \nonumber
\end{align}
We then minimize the divergence between $p_i$ and $y_i$:
\begin{align}
\delta^* &= 
\arg\min_{\delta_i^L}
\Big[
-\log P_{\theta}(o_i\,|\,s_i,r_i;h_L+\delta_i^L)\\
&+\lambda_{\text{KL}}\,\mathrm{KL}(y_i \,\|\, p_i)
+\lambda_{\text{MSE}}\,\|p_i - y_i\|_2^2
\Big],
\label{eq:bit_alignment}
\end{align}
where the first term preserves the model's prediction. We guide the embedding's
cosine similarity distribution toward the one-hot bit assignment by minimizing
the Kullback-Leibler (KL) divergence between $p_i$ and $y_i$. Since the KL
divergence primarily captures distributional differences, we strengthen the
alignment by minimizing the mean squared error (MSE) between $p_i$ and $y_i$.
The hyperparameters $\lambda_{\text{KL}}$ and $\lambda_{\text{MSE}}$ balance
between these two objectives.

\noindent\underline{\emph{Parameter Update.}} After obtaining the optimized
offset $\{\delta_i\}_{i=1}^{N}$ for all anchor samples, we compute their
corresponding key embeddings $\{k_i\}_{i=1}^{N_t}$ from the last subject tokens.
Following the closed-form formulation of ME~\cite{meng2023memit}, the final
parameter update $\Delta_L$ at layer~$L$ is derived by solving the following
linear system:
\begin{equation}
\Delta_L
\!\left(
C_p + 
\sum_{i=1}^{N} k_i k_i^{\top}
\right)
=
\sum_{i=1}^{N} \delta_i^L k_i^{\top},
\label{eq:final_update}
\end{equation}
where $C_p=\lambda\,\mathbb{E}_{k_p}[\,k_p k_p^{\top}\,]$ approximates the
covariance of preserved knowledge, and $\lambda$ controls the balance between
preserving existing knowledge and encoding new bits. The closed-form solution is
then
\begin{equation}
\Delta_L
=
\left(
\sum_{i=1}^{N} \delta_i^L k_i^{\top}
\right)
\!\left(
C_p + 
\sum_{i=1}^{N} k_i k_i^{\top}
\right)^{-1}.
\label{eq:delta_solution}
\end{equation}
Finally, the model parameters are updated as $W_L^{\text{new}} =
W_L^{\text{old}} + \Delta_L$,
where the watermark is encoded within the hidden representation
subspace of the MLP modules. 

The updates can also be distributed across multiple candidate layers to further
enhance robustness and minimize the overall impact on the model. This is
achieved by sequentially updating all selected layers from shallow to deep. For
each layer $\ell$ in all candidate layers, we compute the corresponding update
$\Delta_i^\ell$ following Equation~\ref{eq:delta_solution} by downscaling the
optimized offset $\delta_i^L$ in each layer with $\lambda_\ell$, that is,
$\delta_i^\ell = \delta_i^L/\lambda_\ell$, where $\lambda_\ell$ decreases with
layer depth and equals $1$ for the final layer~\cite{meng2023memit}. We discuss
the details of different layer selection strategies in
\Cref{sec:hyperparameters}. Overall, this global update embeds the optimized
watermark bits into the model's memory while maintaining high fidelity to its
original functionality.

\subsection{White-Box Verification}
\label{sec:whitebox}
As shown in Figure~\ref{fig:seal_methodology} (b), we can directly inspect the
hidden representations of the predefined anchor samples to verify the embedded
watermark bits under white-box access.

\noindent\textbf{Direct Embedding Inspection.} 
For each anchor sample $(s_i, r_i, o_i)$ at the target layer $L$, 
we extract the hidden representation of the last subject token, $h_L(s_i)$, 
and compute its cosine similarities with the two bit vectors 
$\{v_0^{(i)}, v_1^{(i)}\}$ representing the binary codes $\{0,1\}$:
\begin{equation}
\text{cos}_0 = \cos(h_L(s_i), v_0^{(i)}), \qquad
\text{cos}_1 = \cos(h_L(s_i), v_1^{(i)}).
\end{equation}
The recovered bit $\hat{b}_i$ is set to $1$ if $\text{cos}_1 > \text{cos}_0$,
and $0$ otherwise. By performing this comparison across all anchor samples, we
can recover the complete watermark bit sequence $\hat{b} = (\hat{b}_1, \ldots,
\hat{b}_N)$.

\subsection{Black-Box Verification}
\label{sec:blackbox}
In the black-box setting, the defender has no access to the model's internal
parameters or hidden representations and can only query its output logits.  
To verify the embedded watermark under such constraints, we first examine how
each bit vector influences the output logits after watermark insertion. The
resulting logit pattern is then treated as a unique identifier for subsequent
verification. 

\noindent \textbf{Counteract with Adversaries.}~When the model is modified by
fine-tuning, distillation, or other adversarial operations that aim to remove
the watermark, the corresponding logit signatures may drift. To counteract this
effect, we introduce a \emph{Bayesian reanchoring} procedure that adaptively
compensates for model drift and restores the alignment between the bit vectors
and their reference signatures based on precomputed information before the
watermarked model is deployed. Notice that we do not assume the defender has
knowledge of the exact modification process applied to the model. Instead, we
only assume access to a small set of reference triplets that are different from
the anchor samples used during watermark insertion.

We proceed to introduce the details of our black-box verification framework from
three key components: (1) precomputed signatures before deployment,  
(2) Bayesian re-anchoring procedure to mitigate model drift, and  
(3) bit-wise logit comparison for final verification. We demonstrate the
verification process in Figure~\ref{fig:seal_methodology} (c).

\subsubsection{Precomputed Bit Signatures}
\label{sec:blackbox_bit_signature}
Before deploying the watermarked model, we precompute the influence of each bit
vector on the output logits to obtain its corresponding \emph{bit signature}.  
These signatures act as practical approximations of the bit vectors in the logit
space and can serve as reference points for subsequent black-box verification.
To estimate the influence of each bit vector, we collect an auxiliary set of
reference triplets $\mathcal{D}_\text{ref} = \{(s_k, r_k, o_k)\}_{k=1}^K$.  
For each bit vector $v_{b_i}$, we simulate the watermark insertion process by
slightly perturbing the hidden state of the last subject token in
$\mathcal{D}_\text{ref}$ using $v_{b_i}$, and then observe the induced changes
in the output logits. Specifically, at the target layer $L$, the hidden state of
the last subject token is perturbed as
\begin{equation}
h_L'(s_k) = h_L(s_k) + \eta v_{b_i},
\end{equation}
where $\eta$ is a small scaling factor.  
The perturbed hidden state is then projected to the vocabulary logits by the
rest part of the LLM $f(\cdot)$. The influence of $v_{b_i}$ on the output logits
is estimated by averaging the logit differences across all reference triplets:
\begin{align}
&\frac{1}{K}\sum_{k=1}^K \big(f(h_L'(s_k)) - f(h_L(s_k))\big)\\
&= \frac{1}{K}\sum_{k=1}^K \big(f(h_L(s_k)+\eta v_{b_i}) - f(h_L(s_k))\big) \approx f(v_{b_i})
\label{eq:sentinel_projection_blackbox_reorder}
\end{align}
Here, the scaling factor $\eta$ is set sufficiently small such that the
perturbation lies within the locally linear regime of the logit mapping.  
Under this assumption, the change of output logits can be approximated by
a linear response to the injected direction $v_{b_i}$, allowing us to estimate
the influence of each bit vector as $f(v_{b_i})$.

We then select a set of representative tokens that exhibit the highest logit
responses with respect to $v_{b_i}$, referred to as the \emph{sentinel tokens},
denoted by $\mathcal{T}=\{t_1,\ldots,t_m\}$.  
Their projection responses are recorded as
\begin{equation}
w_{\text{sig}}^{(i)} = \{f(v_{b_i})[t_j]|j=1,\ldots,m\}
\end{equation}
where $f(v_{b_i})[t_j]$ denotes the logit of the $j$-th sentinel token for the
bit vector $v_{b_i}$. Each $w_{\text{sig}}^{(i)}$ serves as a unique reference
signature for the $i$-th bit within the logit space, providing a stable basis
for black-box verification in later stages. 

\subsubsection{Black-Box Bayesian Reanchoring}
\label{sec:blackbox_reanchor}
An adversary's attempt to remove the watermark through fine-tuning or
knowledge distillation may cause the bit-vector direction $v_b$ to drift,
thereby shifting its projection on the sentinel logits.  
To correct this drift and recover the effective sentinel weights under the
deployed model, we propose a \emph{Bayesian reanchoring} procedure.  
This process first estimates the logit drift between the original watermarked
model and the deployed model using the same set of reference triplets
$\mathcal{D}_{\text{ref}}$, and then infers the posterior change within the
logit subspace to reconstruct the effective sentinel weights.

\noindent\textbf{Estimating Logit Drift.}  
To estimate the logit drift, we first record the sentinel logits of the
watermarked model $f^{W}(\cdot)$ on the reference dataset
$\mathcal{D}_{\text{ref}}$:
\begin{equation}
\mathcal{L}_k^{\text{W}} = f^{W}(s_k)[\mathcal{T}] = S_k + w_{\text{sig}} + \epsilon_k^0,
\end{equation}
where $S_k$ denotes the prompt-dependent semantic term.  

For the deployed model $f^{D}(\cdot)$, the defender can query it with the same
dataset $\mathcal{D}_{\text{ref}}$ and collect the logits restricted to the
sentinel token set $\mathcal{T}$:
\begin{equation}
\mathcal{L}_k^{\text{D}} = f^{D}(s_k)[\mathcal{T}] = S_k + w^* + \epsilon_k,
\end{equation}
where $w^*\!\in\!\mathbb{R}^{|\mathcal{T}|}$ denotes the latent sentinel-weight
vector.  

To eliminate the semantic component $S_k$, we compute the paired difference
between the two models' logits:
\begin{equation}
\Delta \mathcal{L}_k = \mathcal{L}_k^{\text{D}} - \mathcal{L}_k^{\text{W}}
= (w^* - w_{\text{sig}}) + (\epsilon_k - \epsilon_k^0),
\label{eq:deltaL_blackbox_reorder}
\end{equation}
which yields an unbiased estimate of the drift between the deployed and
reference signatures. We summarize these samples via their empirical mean and
covariance:
\begin{equation}
\mu_\Delta = \mathbb{E}[\Delta \mathcal{L}], 
\qquad 
\Sigma_\Delta = \text{Cov}(\Delta \mathcal{L}),
\end{equation}
which characterize the average drift and the variation within the sentinel logit
subspace.

\noindent\textbf{Bayesian Reanchoring.}  
To correct the observed logit drift and recover the effective sentinel weights
under the deployed model, we perform \emph{Bayesian reanchoring} based on the
precomputed sentinel signatures.

Assuming a Gaussian prior 
$p(w)=\mathcal{N}(w_{\text{sig}}, \lambda^{-1}I)$ 
and a Gaussian likelihood 
$p(\mu_\Delta|w)=\mathcal{N}(w - w_{\text{sig}}, \rho\Sigma_\Delta)$,  
the posterior mean admits the closed-form solution:
\begin{equation}
w^*
= (\rho \Sigma_\Delta + \lambda I)^{-1}
(\,w_{\text{sig}} + P \mu_\Delta + \lambda P w_{\text{sig}}\,),
\label{eq:bayes_reanchor_reorder}
\end{equation}
where $P = I - \frac{\mathbf{1}\mathbf{1}^\top}{m}$ removes uniform offsets from
the estimated directions. The resulting posterior mean $w^*$ denotes the
reanchored sentinel weights under the deployed model, effectively compensating
for model drift. We leave the detailed derivation in the Appendix
\ref{sec:bayes_reanchor_derivation}.

\subsubsection{Bit-Wise Verification}
\label{sec:blackbox_verify_reorder}
With reanchored sentinel weights $\{w_0^{(i)}, w_1^{(i)}\}$, we can verify the
watermark by comparing the logit responses of the anchor samples to the
reanchored sentinel weights.

For each anchor sample $(s_i, r_i, o_i)$, we collect the sentinel logits of the
subject token of $s_i$:
\begin{equation}
\mathcal{L}_i = f^{D}(s_i)[\mathcal{T}] \in \mathbb{R}^{|\mathcal{T}|}.
\end{equation}
We then compute inner products with the two weights:
\begin{equation}
s_0 = \langle \mathcal{L}_i,\, w_{0}^{(i)} \rangle, 
\qquad
s_1 = \langle \mathcal{L}_i,\, w_{1}^{(i)} \rangle,
\end{equation}
and determine the recovered bit $\hat{b}_i$ as $1$ if $s_1 > s_0$, and $0$
otherwise. Repeating this for all anchor samples yields the recovered bit
sequence $\hat{b}=(\hat{b}_1,\ldots,\hat{b}_N)$.  

\subsection{Security and Practicality Explanation}
\subsubsection{Security Assumptions}
Following the above technical pipeline, we now summarize the security assumption
and information visibility of SEAL. By clarifying these, we better explain its
security strength, even in front of advanced, ``knowledgeable'' attackers (see
below).

\noindent\textbf{Defender Side.} The defender privately maintains the following
information: (1) a set of $N$ anchor samples $\mathcal{A} = \{(s_i, r_i,
o_i)\}_{i=1}^N$, which serve as secret semantic carriers for watermark
embedding; (2) a corresponding set of $2N$ orthogonal bit vectors $\mathcal{V} =
\{v_0^{(i)}, v_1^{(i)}\}_{i=1}^N$ that define the subspace encoding of each bit;
and (3) for black-box verification purposes, notice a sentinel token set
$\mathcal{T}$ and reference dataset $\mathcal{D}_\text{ref}$ for Bayesian
reanchoring are generated prior to deployment; these are also kept secret.
These elements collectively constitute the defender's ``secret key'' and are
neither disclosed nor inferable by adversaries.

\noindent\textbf{Public Information.} The embedding and verification algorithms
of SEAL, including the training objectives, loss functions, and reanchoring
procedures, are \textit{public} by design. However, adversaries do not have
access to the exact anchor samples or bit vectors that define the encoded
subspace.

\noindent\textbf{Adversary Knowledge.} Although the SEAL procedure is publicly
available, \underline{ordinary attackers} \emph{do not} have access to the
specific configuration, such as the exact layers for watermark embedding. This
is reasonable, as such implementation details are rarely disclosed in practice,
and modern LLMs often have complex architectures with numerous layers, making
exhaustive guessing impractical. Given that, we also consider more
\underline{knowledgeable attackers} who are aware of the general SEAL mechanism
and may even infer the target layer used for watermark injection. This
represents a realistic scenario in which adversaries can deduce partial
implementation details from public information. Based on the results
(Section~\ref{sec:hyperparameters}), different layers manifest distinct support
on watermarking effectiveness. Thus, we assume that knowledgeable, advanced
attackers may identify certain layers that have a high chance of containing
watermarks (e.g., those “deep” layers). We name these attackers as
“knowledgeable attackers” and assess them in Section~\ref{sec:adaptive_attacks}. 

\subsubsection{Usage in Practice}
We now discuss using SEAL in practice for lineage identification, ownership
verification, and model deployment.

\noindent\textbf{Lineage Identification.} Model lineage is identified by
comparing the cosine similarity between the recovered bit sequences of two
models. A high similarity score indicates that the two models are derived from
the same watermarked source, without requiring an exact bit-wise match.

\noindent\textbf{Ownership Verification.} After recovering $\hat{b} \in
\{0,1\}^N$ using either white-box or black-box verification methods, the
defender can compare $\hat{b}$ with the registered signature $b^*$ and compute
the \emph{bit error rate (BER)} as
\begin{equation}
\mathrm{BER} = \frac{1}{N} \sum_{i=1}^N \mathbb{I}[\hat{b}_i \neq b^*_i].
\end{equation}
Ownership is claimed when $\mathrm{BER} < \tau$, where $\tau$ is a predefined
verification threshold. A smaller $\tau$ leads to a negligible false positive
rate (FPR) in ownership claims, since the probability that a random watermark
coincidentally matches the defender's signature decreases \textit{exponentially}
with $N$. For example, when $\tau = 0$, a 64-bit watermark yields an FPR of
$2^{-64}$, which is practically zero for IP-related applications. Therefore, in
IP protection scenarios (which are often high-stake and legally binding), we
conservatively adopt an \emph{exact-match} criterion (i.e., $\tau = 0$) to avoid
ambiguity in ownership claims. Our experiments in
Section~\ref{sec:watermark_extraction} demonstrate that SEAL achieves near-zero
BER, enabling deterministic verification without relying on probabilistic
thresholds.

In practice, bit errors may occur due to statistical randomness or even hardware
faults. To mitigate this, we further discuss incorporating error-correcting
codes (ECC)~\cite{hamming1950error, bose1960class, wicker1999reed} into the
watermark encoding to tolerate minor bit corruption in
Appendix~\ref{sec:ecc}.

\noindent\textbf{Deploying SEAL.} To deploy SEAL, the defender generates and
stores $2N$ bit vectors and their corresponding anchor set for each model
instance prior to release. This one-time cost is negligible compared to model
pretraining or fine-tuning. During verification, SEAL only requires access to
hidden representations (for white-box verification) or output logits (for
black-box verification). Consequently, the framework is fully compatible with
API-based verification, where access to model logits is commonly available in
many LLM APIs~\cite{hills2023using_logprobs, li2025differentiationbasedextractionproprietarydata}, and introduces minimal
computational overhead.

%% file: docs/05-evaluation.tex
\section{Evaluation Setup and Configuration}
\noindent\textbf{Evaluation Setup.} We evaluate our method on six representative
LLMs: LLaMA2-7B~\cite{touvron2023llama}, LLaMA3-8B~\cite{dubey2024llama},
Qwen2.5-7B~\cite{qwen2025qwen25technicalreport},
Qwen2.5-14B~\cite{qwen2025qwen25technicalreport},
Mistral-7B~\cite{jiang2023mistral7b}, and
DeepSeek-Chat-7B~\cite{liu2024deepseek}. To assess robustness, we consider five
categories of model-level attacks that aim to remove the embedded watermark or
disrupt fingerprint extraction: SFT~\cite{xu-etal-2024-instructional},
PEFT~\cite{hu2022lora, wang2024lora},
DT~\cite{gu2024minillmknowledgedistillationlarge, peng-etal-2025-pre,
zhang-etal-2024-plad}, QT~\cite{xiao2023smoothquant, banner2019post,
lin2024awq}, and MM~\cite{yang2024model, li2025model,
nobari2025activationinformed}.

For SFT attacks, we perform full-parameter fine-tuning on two public
instruction-following datasets, Alpaca~\cite{taori2023stanford} and
UltraChat~\cite{ding-etal-2023-enhancing}. For PEFT attacks, we inject LoRA
adapters into the model and fine-tune only the inserted low-rank parameters. For
distillation attacks, we assume a setting where the watermark is injected into a
student model while the adversary distills knowledge from a larger or fine-tuned
teacher model into the watermarked student \cite{jiang-etal-2023-lion,
zhang-etal-2024-dual, li-etal-2025-learning}. 

Notice that, based on our experiments, modification on those deeper layers
manifests poorer influence on the output logits. Thus, as a reasonable
assumption, we consider attackers tend to target those ``shallow'' layers for
modification to maximize the impact on model behavior. Hence, without loss of
generality, for these attacks, we assume the attacker targets layers
(5, 6, 7, 8, 9, 10) by default. We further discuss the impact of this assumption in
Section~\ref{sec:adaptive_attacks}.

To simulate quantization, we apply symmetric per-tensor fake quantization, where
each floating-point weight tensor is scaled by its maximum absolute value and
projected onto a uniform $n$-bit integer grid before being dequantized back to
floating point. We set $n=8$ in all experiments \cite{jacob2018quantization}.
Finally, for model-merging attacks, we linearly interpolate the parameters of
the watermarked model and its fine-tuned counterpart for the merged model
$\theta'_{\text{merged}} = \alpha\,\theta'_{\text{watermarked}} + (1 -
\alpha)\,\theta'_{\text{fine-tuned}}$ where $\alpha = 0.7$ in all experiments.

\noindent\textbf{Baseline Methods.} We compare our approach against a
comprehensive set of existing fingerprinting and watermarking methods. Among
them, Gradient~\cite{wu2025gradient, shao2025sok}, Huref~\cite{zeng2024huref},
Reef~\cite{zhang2024reefrepresentationencodingfingerprints}, and
PDF~\cite{yoon2025intrinsicfingerprintllmscontinue} require \textit{white-box}
access to the model parameters. In contrast, LLMMap~\cite{llmmap},
MET~\cite{gao2025model}, SEF~\cite{shao2025sok}, and
Trap~\cite{gubri-etal-2024-trap} operate under the \textit{black-box} setting,
where only model outputs or logits are accessible. We further include
backdoor-based watermarking baselines, namely SFP~\cite{nasery2025scalable} and
IF~\cite{xu-etal-2024-instructional}. For IF, we evaluate two variants: IF-sft,
which fine-tunes transformer layers using full-parameter SFT, and IF-emb, which
fine-tunes only the embedding layer.

\begin{table*}[htbp]
  \centering
  \caption{Model lineage identification performance of SEAL and baselines under
  white-box and black-box settings across six LLMs. SEAL(W) and SEAL(B) denote
  the white-box and black-box variants.}
  \setlength{\tabcolsep}{2.5pt}
  \resizebox{0.95\textwidth}{!}{%
  \begin{tabular}{ccccccccccccccc}
  \toprule
  \multirow{2}{*}{\textbf{Model}} & \multirow{2}{*}{\textbf{Metric}} & \multicolumn{5}{c}{\textbf{White-box}} & \multicolumn{8}{c}{\textbf{Black-box}} \\
  \cmidrule(lr){3-7} \cmidrule(lr){8-15}
  & & \textbf{Gradient} & \textbf{Huref} & \textbf{Reef} & \textbf{PDF} & \textbf{SEAL(W)} & \textbf{LLMMap} & \textbf{MET} & \textbf{SEF} & \textbf{Trap} & \textbf{SFP} & \textbf{IF-sft} & \textbf{IF-emb} & \textbf{SEAL(B)} \\
  \midrule
  \multirow{3}{*}{\textbf{LLaMA2-7B}} 
  & AUC$\uparrow$ & 0.53\scriptsize $\pm$0.01 & 1.00\scriptsize $\pm$0.00 & 1.00\scriptsize $\pm$0.02 & 1.00\scriptsize $\pm$0.00 & \textbf{1.00}\scriptsize $\pm$0.00 & 0.30\scriptsize $\pm$0.01 & 0.47\scriptsize $\pm$0.01 & 0.28\scriptsize $\pm$0.01 & 0.44\scriptsize $\pm$0.03 & 0.67\scriptsize $\pm$0.04 & 0.50\scriptsize $\pm$0.00 & 0.50\scriptsize $\pm$0.00 & \textbf{1.00}\scriptsize $\pm$0.00 \\
  & pAUC$\uparrow$ & 0.33\scriptsize $\pm$0.01 & 1.00\scriptsize $\pm$0.00 & 1.00\scriptsize $\pm$0.02 & 1.00\scriptsize $\pm$0.00 & \textbf{1.00}\scriptsize $\pm$0.00 & 0.17\scriptsize $\pm$0.01 & 0.00\scriptsize $\pm$0.00 & 0.00\scriptsize $\pm$0.00 & 0.33\scriptsize $\pm$0.01 & 0.33\scriptsize $\pm$0.03 & 0.00\scriptsize $\pm$0.00 & 0.00\scriptsize $\pm$0.00 & \textbf{1.00}\scriptsize $\pm$0.00 \\
  & MD$\uparrow$ & 0.25\scriptsize $\pm$0.03 & 1.84\scriptsize $\pm$0.00 & 1.91\scriptsize $\pm$0.06 & 1.52\scriptsize $\pm$0.00 & \textbf{1.92}\scriptsize $\pm$0.01 & 0.70\scriptsize $\pm$0.02 & 0.87\scriptsize $\pm$0.04 & 1.34\scriptsize $\pm$0.01 & 1.41\scriptsize $\pm$0.07 & 0.86\scriptsize $\pm$0.04 & 0.00\scriptsize $\pm$0.00 & 0.00\scriptsize $\pm$0.00 & \textbf{1.89}\scriptsize $\pm$0.02 \\
  \midrule
  \multirow{3}{*}{\textbf{Qwen2.5-7B}} 
  & AUC$\uparrow$ & 0.97\scriptsize $\pm$0.00 & 1.00\scriptsize $\pm$0.00 & 1.00\scriptsize $\pm$0.02 & 1.00\scriptsize $\pm$0.00 & \textbf{1.00}\scriptsize $\pm$0.00 & 0.67\scriptsize $\pm$0.02 & 0.18\scriptsize $\pm$0.01 & 0.67\scriptsize $\pm$0.03 & 0.50\scriptsize $\pm$0.02 & 0.50\scriptsize $\pm$0.02 & 0.50\scriptsize $\pm$0.00 & 0.50\scriptsize $\pm$0.00 & \textbf{1.00}\scriptsize $\pm$0.00 \\
  & pAUC$\uparrow$ & 0.83\scriptsize $\pm$0.01 & 1.00\scriptsize $\pm$0.00 & 1.00\scriptsize $\pm$0.01 & 1.00\scriptsize $\pm$0.00 & \textbf{1.00}\scriptsize $\pm$0.00 & 0.67\scriptsize $\pm$0.01 & 0.00\scriptsize $\pm$0.00 & 0.67\scriptsize $\pm$0.02 & 0.00\scriptsize $\pm$0.00 & 0.00\scriptsize $\pm$0.00 & 0.00\scriptsize $\pm$0.00 & 0.00\scriptsize $\pm$0.00 & \textbf{1.00}\scriptsize $\pm$0.00 \\
  & MD$\uparrow$ & 1.54\scriptsize $\pm$0.02 & 1.91\scriptsize $\pm$0.00 & 1.71\scriptsize $\pm$0.05 & 1.79\scriptsize $\pm$0.00 & \textbf{1.92}\scriptsize $\pm$0.02 & 0.15\scriptsize $\pm$0.01 & 1.34\scriptsize $\pm$0.03 & 1.50\scriptsize $\pm$0.04 & 0.00\scriptsize $\pm$0.00 & 0.00\scriptsize $\pm$0.00 & 0.00\scriptsize $\pm$0.00 & 0.00\scriptsize $\pm$0.00 & \textbf{1.91}\scriptsize $\pm$0.01 \\
  \midrule
  \multirow{3}{*}{\textbf{LLaMA3-8B}} 
  & AUC$\uparrow$ & 0.94\scriptsize $\pm$0.02 & 1.00\scriptsize $\pm$0.00 & 1.00\scriptsize $\pm$0.02 & 0.94\scriptsize $\pm$0.00 & \textbf{1.00}\scriptsize $\pm$0.00 & 0.33\scriptsize $\pm$0.01 & 0.67\scriptsize $\pm$0.02 & 0.28\scriptsize $\pm$0.02 & 0.33\scriptsize $\pm$0.02 & 0.58\scriptsize $\pm$0.02 & 0.50\scriptsize $\pm$0.00 & 0.50\scriptsize $\pm$0.00 & \textbf{1.00}\scriptsize $\pm$0.00 \\
  & pAUC$\uparrow$ & 0.67\scriptsize $\pm$0.01 & 1.00\scriptsize $\pm$0.00 & 1.00\scriptsize $\pm$0.02 & 0.67\scriptsize $\pm$0.00 & \textbf{1.00}\scriptsize $\pm$0.00 & 0.33\scriptsize $\pm$0.02 & 0.33\scriptsize $\pm$0.01 & 0.00\scriptsize $\pm$0.00 & 0.33\scriptsize $\pm$0.01 & 0.17\scriptsize $\pm$0.01 & 0.00\scriptsize $\pm$0.00 & 0.00\scriptsize $\pm$0.00 & \textbf{1.00}\scriptsize $\pm$0.00 \\
  & MD$\uparrow$ & 1.51\scriptsize $\pm$0.00 & 1.90\scriptsize $\pm$0.00 & 1.85\scriptsize $\pm$0.06 & 1.52\scriptsize $\pm$0.00 & \textbf{1.91}\scriptsize $\pm$0.02 & 0.58\scriptsize $\pm$0.03 & 0.86\scriptsize $\pm$0.03 & 1.43\scriptsize $\pm$0.03 & 1.90\scriptsize $\pm$0.04 & 0.58\scriptsize $\pm$0.03 & 0.00\scriptsize $\pm$0.00 & 0.00\scriptsize $\pm$0.00 & \textbf{1.92}\scriptsize $\pm$0.01 \\
  \midrule
  \multirow{3}{*}{\textbf{Mistral-7B}} 
  & AUC$\uparrow$ & 1.00\scriptsize $\pm$0.01 & 1.00\scriptsize $\pm$0.00 & 1.00\scriptsize $\pm$0.02 & 1.00\scriptsize $\pm$0.00 & \textbf{1.00}\scriptsize $\pm$0.00 & 0.00\scriptsize $\pm$0.00 & 0.25\scriptsize $\pm$0.01 & 0.00\scriptsize $\pm$0.00 & 0.42\scriptsize $\pm$0.02 & 0.58\scriptsize $\pm$0.02 & 0.50\scriptsize $\pm$0.00 & 0.50\scriptsize $\pm$0.00 & \textbf{1.00}\scriptsize $\pm$0.00 \\
  & pAUC$\uparrow$ & 1.00\scriptsize $\pm$0.00 & 1.00\scriptsize $\pm$0.00 & 1.00\scriptsize $\pm$0.01 & 1.00\scriptsize $\pm$0.00 & \textbf{1.00}\scriptsize $\pm$0.00 & 0.00\scriptsize $\pm$0.00 & 0.00\scriptsize $\pm$0.00 & 0.00\scriptsize $\pm$0.00 & 0.00\scriptsize $\pm$0.00 & 0.17\scriptsize $\pm$0.02 & 0.00\scriptsize $\pm$0.00 & 0.00\scriptsize $\pm$0.00 & \textbf{1.00}\scriptsize $\pm$0.00 \\
  & MD$\uparrow$ & 1.58\scriptsize $\pm$0.00 & 1.91\scriptsize $\pm$0.00 & 1.87\scriptsize $\pm$0.04 & 1.76\scriptsize $\pm$0.00 & \textbf{1.94}\scriptsize $\pm$0.01 & 1.55\scriptsize $\pm$0.01 & 1.11\scriptsize $\pm$0.03 & 1.90\scriptsize $\pm$0.02 & 0.57\scriptsize $\pm$0.03 & 0.58\scriptsize $\pm$0.04 & 0.00\scriptsize $\pm$0.00 & 0.00\scriptsize $\pm$0.00 & \textbf{1.90}\scriptsize $\pm$0.00 \\
  \midrule
  \multirow{3}{*}{\textbf{Deepseek-7B}} 
  & AUC$\uparrow$ & 0.80\scriptsize $\pm$0.01 & 1.00\scriptsize $\pm$0.00 & 1.00\scriptsize $\pm$0.02 & 1.00\scriptsize $\pm$0.00 & \textbf{1.00}\scriptsize $\pm$0.00 & 0.00\scriptsize $\pm$0.00 & 0.28\scriptsize $\pm$0.01 & 0.00\scriptsize $\pm$0.00 & 0.50\scriptsize $\pm$0.03 & 0.50\scriptsize $\pm$0.02 & 0.50\scriptsize $\pm$0.00 & 0.50\scriptsize $\pm$0.00 & \textbf{1.00}\scriptsize $\pm$0.00 \\
  & pAUC$\uparrow$ & 0.50\scriptsize $\pm$0.02 & 1.00\scriptsize $\pm$0.00 & 1.00\scriptsize $\pm$0.01 & 1.00\scriptsize $\pm$0.00 & \textbf{1.00}\scriptsize $\pm$0.00 & 0.00\scriptsize $\pm$0.00 & 0.00\scriptsize $\pm$0.00 & 0.00\scriptsize $\pm$0.00 & 0.00\scriptsize $\pm$0.00 & 0.00\scriptsize $\pm$0.00 & 0.00\scriptsize $\pm$0.00 & 0.00\scriptsize $\pm$0.00 & \textbf{1.00}\scriptsize $\pm$0.00 \\
  & MD$\uparrow$ & 0.97\scriptsize $\pm$0.00 & 1.88\scriptsize $\pm$0.00 & 1.91\scriptsize $\pm$0.07 & 1.68\scriptsize $\pm$0.00 & \textbf{1.91}\scriptsize $\pm$0.01 & 1.52\scriptsize $\pm$0.04 & 1.55\scriptsize $\pm$0.03 & 1.84\scriptsize $\pm$0.02 & 0.00\scriptsize $\pm$0.00 & 0.00\scriptsize $\pm$0.00 & 0.00\scriptsize $\pm$0.00 & 0.00\scriptsize $\pm$0.00 & \textbf{1.89}\scriptsize $\pm$0.01 \\
  \midrule
  \multirow{3}{*}{\textbf{Qwen2.5-14B}} 
  & AUC$\uparrow$ & 1.00\scriptsize $\pm$0.00 & 1.00\scriptsize $\pm$0.00 & 1.00\scriptsize $\pm$0.00 & 1.00\scriptsize $\pm$0.00 & \textbf{1.00}\scriptsize $\pm$0.00 & 0.31\scriptsize $\pm$0.02 & 0.42\scriptsize $\pm$0.02 & 0.50\scriptsize $\pm$0.03 & 0.28\scriptsize $\pm$0.02 & 0.50\scriptsize $\pm$0.01 & 0.50\scriptsize $\pm$0.00 & 0.50\scriptsize $\pm$0.00 & \textbf{1.00}\scriptsize $\pm$0.00 \\
  & pAUC$\uparrow$ & 1.00\scriptsize $\pm$0.01 & 1.00\scriptsize $\pm$0.00 & 1.00\scriptsize $\pm$0.00 & 1.00\scriptsize $\pm$0.00 & \textbf{1.00}\scriptsize $\pm$0.00 & 0.17\scriptsize $\pm$0.01 & 0.00\scriptsize $\pm$0.00 & 0.50\scriptsize $\pm$0.02 & 0.00\scriptsize $\pm$0.00 & 0.00\scriptsize $\pm$0.00 & 0.00\scriptsize $\pm$0.00 & 0.00\scriptsize $\pm$0.00 & \textbf{1.00}\scriptsize $\pm$0.00 \\
  & MD$\uparrow$ & 1.53\scriptsize $\pm$0.02 & 1.88\scriptsize $\pm$0.00 & 1.90\scriptsize $\pm$0.00 & 1.65\scriptsize $\pm$0.00 & \textbf{1.92}\scriptsize $\pm$0.01 & 0.53\scriptsize $\pm$0.02 & 0.58\scriptsize $\pm$0.02 & 0.39\scriptsize $\pm$0.03 & 1.42\scriptsize $\pm$0.03 & 0.00\scriptsize $\pm$0.00 & 0.00\scriptsize $\pm$0.00 & 0.00\scriptsize $\pm$0.00 & \textbf{1.89}\scriptsize $\pm$0.00 \\
  \bottomrule
  \end{tabular}%
  }
  \label{tab:rq1_copyright}
\end{table*}

\noindent\textbf{Evaluation Metrics.} We first evaluate the effectiveness of
identifying model lineage. For fair comparison with existing fingerprinting
methods, we adopt the same evaluation metrics (AUC, pAUC, and MD) commonly used in prior literature
\cite{shao2025sok}. A higher AUC, pAUC, or MD indicates stronger discriminative
power. We refer to Appendix~\ref{sec:metric_detail_li} for more details.
We then compute these metrics for each model when
compared against others, as summarized in Table~\ref{tab:comparepool}.

To further evaluate the characteristics of our watermarking approach, we design
the following metrics to assess the multi-bit watermarking capacity:
\begin{itemize}[leftmargin=1em]
    \item \textbf{Bit Error Rate (BER).} The proportion of incorrectly recovered
    bits relative to the originally embedded bits. Lower BER denotes more
    reliable watermark extraction. We report all BER values as percentages (\%).
    
    \item \textbf{Bit Separability (BSE).} The average gap between a watermarked
    bit vector’s similarity to its corresponding versus non-corresponding
    original bit vectors. Higher BSE implies stronger bit separability.
    
    \item \textbf{Bit Stability (BST).} The standard deviation of the similarity
    scores between watermarked bits and their original counterparts across
    repeated extractions. Lower BST implies greater robustness and stability
    of the watermark.
\end{itemize}

\noindent\textbf{SEAL Settings.} We embed multi-bit watermarks into the model
and evaluate under both \textit{white-box} and \textit{black-box} settings. By
default, each model is embedded with a $64$-bit watermark. We also extend the
evaluation to different bit lengths, including 32, 64, 128, 256, 512, and
1024 bits. For watermark embedding, we select samples from the
\textit{CounterFact} dataset~\cite{rome}. Specifically, we select samples on
which the model exhibits high prediction confidence, enabling a more stable
optimization during watermark embedding. We further discuss this choice in
Section~\ref{sec:hyperparameters}.

\section{Evaluation}
\label{sec:evaluation}
We organize our evaluation around the following five research questions (RQs):
\textbf{RQ1:} How effective is SEAL in identifying model lineage compared with
existing methods? \textbf{RQ2:} How effective is SEAL in extracting watermarks
under various attack behaviors? \textbf{RQ3:} How harmless is SEAL to the
model's original performance and utility? \textbf{RQ4:} How does SEAL perform in
watermark extraction across different bit lengths, in terms of time efficiency
and GPU usage? \textbf{RQ5:} How do different hyperparameter settings affect the
effectiveness of SEAL in watermark extraction?

\subsection{RQ1: Model Lineage Identification} 
We first evaluate the \textbf{Effectiveness} of SEAL in identifying
derivative models from independent ones, in comparison with existing
fingerprinting methods.

\noindent\textbf{SEAL Effectively Identifies Model Lineage Across Different
LLMs.} As summarized in Table~\ref{tab:rq1_copyright}, SEAL consistently
achieves superior performance in distinguishing derivative models from
independent ones. Across all evaluation settings, SEAL attains perfect AUC and
pAUC scores of 1.00, and also achieves the highest MD across all LLMs,
indicating that the extracted features are highly separable for different model
lineages. These results demonstrate that SEAL achieves state-of-the-art
effectiveness in model lineage identification.

\noindent\textbf{SEAL Maintains High Lineage Identification Accuracy in Both
White-Box and Black-Box Scenarios.} As shown in Table~\ref{tab:rq1_copyright},
SEAL remains highly effective in identifying model lineage even under the
black-box setting, where only query access or output logits are available.
Across all six LLMs, SEAL achieves perfect AUC and pAUC scores of 1.00 when
distinguishing derivative models from independent ones. In contrast, most
existing fingerprinting methods fail to effectively identify model lineage,
performing close to random guessing with AUC values around 0.5. Similarly,
backdoor-based watermarking methods such as SFP and IF also fail to maintain
accuracy, as their injected backdoor behaviors are easily removed through
fine-tuning, rendering their lineage identification capabilities fragile.

In the white-box setting, where full model parameters are accessible, existing
fingerprinting methods demonstrate promising performance in lineage
identification. Among them, Huref, Reef, and PDF achieve near-perfect AUC and
pAUC scores close to 1.00, while Gradient exhibits limited adaptability across
different LLMs, achieving only 0.53 AUC in distinguishing derivative models of
LLaMA2-7B from other independent models. Despite the competitive performance of
these methods, SEAL consistently maintains perfect AUC and pAUC scores of 1.00,
along with the highest MD across all LLMs, indicating superior separability
between derivative and independent models.

Overall, these results highlight SEAL's superior effectiveness in model lineage
identification, particularly under black-box conditions where existing methods
fail to generalize or retain reliability.

\subsection{RQ2: Multi-Bit Watermark Extraction}
\label{sec:watermark_extraction}
Existing fingerprinting and watermarking methods are usually limited to
identifying model lineage or embedding a single-bit piece of information. This
limitation constrains their applicability for sophisticated multi-bit ownership
verification, which requires embedding and extracting multi-bit information. We
therefore evaluate the \textbf{Effectiveness} of SEAL in extracting multi-bit
watermarks under various attack scenarios.

\begin{table}[htbp]
  \centering
  \caption{Multi-bit watermark extraction performance of SEAL under different attack behaviors. }
  \begingroup\setlength{\tabcolsep}{2pt}
  \setlength{\tabcolsep}{2.5pt}
  \resizebox{0.92\columnwidth}{!}{
    \begin{tabular}{@{}ccccccccc@{}}
    \toprule
    \textbf{Evaluation} & \textbf{Bits} & \textbf{Metrics} & \textbf{No attack} & \textbf{SFT} & \textbf{PEFT} & \textbf{DT} & \textbf{QT} & \textbf{MM} \\
    \midrule
    \multirow{12}[7]{*}{\textbf{White-box}} & \multirow{3}[2]{*}{64} & BER$\downarrow$ & 0.06\scriptsize$\pm$1.25 & \textbf{0.00}\scriptsize$\pm$0.00 & 0.63\scriptsize$\pm$1.25 & 0.06\scriptsize$\pm$1.25 & \textbf{0.00}\scriptsize$\pm$0.00 & \textbf{0.00}\scriptsize$\pm$0.00 \\
          &       & BSE$\uparrow$ & 0.04\scriptsize$\pm$0.09 & 0.01\scriptsize$\pm$0.03 & 0.03\scriptsize$\pm$0.09 & 0.05\scriptsize$\pm$0.08 & 0.05\scriptsize$\pm$0.10 & 0.02\scriptsize$\pm$0.06 \\
          &       & BST$\downarrow$ & 0.76\scriptsize$\pm$0.05 & 0.49\scriptsize$\pm$0.16 & 0.67\scriptsize$\pm$0.03 & 0.62\scriptsize$\pm$0.13 & 0.72\scriptsize$\pm$0.05 & 0.38\scriptsize$\pm$0.11 \\
\cmidrule{2-9}          & \multirow{3}[2]{*}{128} & BER$\downarrow$ & 0.00\scriptsize$\pm$0.00 & 0.52\scriptsize$\pm$1.36 & \textbf{0.00}\scriptsize$\pm$0.00 & 0.07\scriptsize$\pm$0.21 & \textbf{0.00}\scriptsize$\pm$0.00 & 0.95\scriptsize$\pm$0.42 \\
          &       & BSE$\uparrow$ & 0.02\scriptsize$\pm$0.06 & 0.01\scriptsize$\pm$0.06 & 0.02\scriptsize$\pm$0.07 & 0.00\scriptsize$\pm$0.05 & 0.03\scriptsize$\pm$0.09 & 0.02\scriptsize$\pm$0.06 \\
          &       & BST$\downarrow$ & 0.75\scriptsize$\pm$0.05 & 0.50\scriptsize$\pm$0.19 & 0.68\scriptsize$\pm$0.02 & 0.56\scriptsize$\pm$0.15 & 0.72\scriptsize$\pm$0.05 & 0.39\scriptsize$\pm$0.11 \\
\cmidrule{2-9}          & \multirow{3}[1]{*}{256} & BER$\downarrow$ & 0.00\scriptsize$\pm$0.00 & 0.15\scriptsize$\pm$0.27 & \textbf{0.00}\scriptsize$\pm$0.00 & 0.10\scriptsize$\pm$0.17 & 0.36\scriptsize$\pm$0.40 & 0.39\scriptsize$\pm$0.55 \\
          &       & BSE$\uparrow$ & 0.02\scriptsize$\pm$0.02 & 0.01\scriptsize$\pm$0.01 & 0.02\scriptsize$\pm$0.01 & 0.01\scriptsize$\pm$0.01 & 0.04\scriptsize$\pm$0.01 & 0.02\scriptsize$\pm$0.01 \\
          &       & BST$\downarrow$ & 0.75\scriptsize$\pm$0.06 & 0.43\scriptsize$\pm$0.21 & 0.60\scriptsize$\pm$0.12 & 0.53\scriptsize$\pm$0.19 & 0.68\scriptsize$\pm$0.03 & 0.27\scriptsize$\pm$0.10 \\
\cmidrule{2-9}          & \multirow{3}[0]{*}{512} & BER$\downarrow$ & 0.08\scriptsize$\pm$0.15 & 0.16\scriptsize$\pm$0.09 & \textbf{0.08}\scriptsize$\pm$0.16 & 0.12\scriptsize$\pm$0.16 & 0.09\scriptsize$\pm$0.16 & 0.15\scriptsize$\pm$0.25 \\
          &       & BSE$\uparrow$ & 0.02\scriptsize$\pm$0.02 & 0.01\scriptsize$\pm$0.01 & 0.02\scriptsize$\pm$0.03 & 0.01\scriptsize$\pm$0.01 & 0.01\scriptsize$\pm$0.03 & 0.02\scriptsize$\pm$0.01 \\
          &       & BST$\downarrow$ & 0.75\scriptsize$\pm$0.04 & 0.46\scriptsize$\pm$0.15 & 0.64\scriptsize$\pm$0.07 & 0.54\scriptsize$\pm$0.23 & 0.70\scriptsize$\pm$0.03 & 0.38\scriptsize$\pm$0.13 \\ 
  \midrule
    \multirow{12}[7]{*}{\textbf{Black-box}}  & \multirow{3}[2]{*}{64} & BER$\downarrow$ & 0.31\scriptsize$\pm$0.63 & 1.03\scriptsize$\pm$1.52 & \textbf{0.16}\scriptsize$\pm$0.47 & 1.09\scriptsize$\pm$2.80 & 0.26\scriptsize$\pm$0.58 & 2.08\scriptsize$\pm$4.00 \\
          &       & BSE$\uparrow$ & 0.02\scriptsize$\pm$0.01 & 0.01\scriptsize$\pm$0.01 & 0.01\scriptsize$\pm$0.01 & 0.01\scriptsize$\pm$0.01 & 0.00\scriptsize$\pm$0.00 & 0.01\scriptsize$\pm$0.01 \\
          &       & BST$\downarrow$ & 0.00\scriptsize$\pm$0.00 & 0.00\scriptsize$\pm$0.00 & 0.00\scriptsize$\pm$0.00 & 0.00\scriptsize$\pm$0.00 & 0.00\scriptsize$\pm$0.00 & 0.00\scriptsize$\pm$0.00 \\
\cmidrule{2-9}          & \multirow{3}[2]{*}{128} & BER$\downarrow$ & 0.00\scriptsize$\pm$0.00 & 1.13\scriptsize$\pm$2.81 & \textbf{0.08}\scriptsize$\pm$0.23 & 1.34\scriptsize$\pm$1.49 & 0.16\scriptsize$\pm$0.31 & 4.06\scriptsize$\pm$3.72 \\
          &       & BSE$\uparrow$ & 0.02\scriptsize$\pm$0.01 & 0.01\scriptsize$\pm$0.01 & 0.01\scriptsize$\pm$0.01 & 0.02\scriptsize$\pm$0.01 & 0.01\scriptsize$\pm$0.01 & 0.01\scriptsize$\pm$0.01 \\
          &       & BST$\downarrow$ & 0.00\scriptsize$\pm$0.00 & 0.00\scriptsize$\pm$0.00 & 0.00\scriptsize$\pm$0.00 & 0.00\scriptsize$\pm$0.00 & 0.00\scriptsize$\pm$0.00 & 0.00\scriptsize$\pm$0.00 \\
\cmidrule{2-9}          & \multirow{3}[2]{*}{256} & BER$\downarrow$ & 0.09\scriptsize$\pm$0.16 & 1.51\scriptsize$\pm$3.95 & 1.21\scriptsize$\pm$1.09 & 1.53\scriptsize$\pm$2.17 & \textbf{0.19}\scriptsize$\pm$0.33 & 4.10\scriptsize$\pm$2.35 \\
          &       & BSE$\uparrow$ & 0.01\scriptsize$\pm$0.00 & 0.01\scriptsize$\pm$0.00 & 0.02\scriptsize$\pm$0.01 & 0.02\scriptsize$\pm$0.01 & 0.01\scriptsize$\pm$0.01 & 0.02\scriptsize$\pm$0.02 \\
          &       & BST$\downarrow$ & 0.00\scriptsize$\pm$0.00 & 0.00\scriptsize$\pm$0.00 & 0.01\scriptsize$\pm$0.00 & 0.01\scriptsize$\pm$0.01 & 0.00\scriptsize$\pm$0.00 & 0.01\scriptsize$\pm$0.00 \\
\cmidrule{2-9}          & \multirow{3}[0]{*}{512} & BER$\downarrow$ & 0.04\scriptsize$\pm$0.08 & 1.68\scriptsize$\pm$2.27 & 1.26\scriptsize$\pm$1.21 & 2.65\scriptsize$\pm$3.27 & \textbf{0.24}\scriptsize$\pm$0.25 & 2.75\scriptsize$\pm$2.38 \\
          &       & BSE$\uparrow$ & 0.02\scriptsize$\pm$0.02 & 0.02\scriptsize$\pm$0.01 & 0.02\scriptsize$\pm$0.01 & 0.02\scriptsize$\pm$0.02 & 0.02\scriptsize$\pm$0.01 & 0.02\scriptsize$\pm$0.02 \\
          &       & BST$\downarrow$ & 0.00\scriptsize$\pm$0.00 & 0.00\scriptsize$\pm$0.00 & 0.00\scriptsize$\pm$0.00 & 0.00\scriptsize$\pm$0.00 & 0.00\scriptsize$\pm$0.00 & 0.00\scriptsize$\pm$0.00 \\ 
    \bottomrule
    \end{tabular}%
  }
  \label{tab:rq2}%
  \endgroup
\end{table}%

\noindent\textbf{SEAL Effectively Extracts Multi-Bit Watermarks Under Various
Attack Behaviors.} As shown in Table~\ref{tab:rq2}, SEAL effectively extracts
watermarks of different bit lengths under all evaluated attack behaviors that
attempt to remove the embedded signals. Across all attack methods and bit
lengths, SEAL achieves an average BER of 0.69\%. Specifically, under SFT
attacks, the error rate of the extracted watermarks remains below 1.51\% across
all configurations. SEAL demonstrates even stronger robustness against PEFT
attacks, where the maximum BER is 1.21\% when extracting 256-bit watermarks in
the black-box setting, while the BER in all other cases remains close to zero.
This is because PEFT updates cause slight changes to the model's parameters
compared to SFT updates. Quantization attacks have a negligible effect on the
alignment between anchor samples and their corresponding bit vectors, with the
highest observed BER of only 0.39\%. Among all evaluated attacks, model merging
poses the greatest challenge to extraction accuracy, as it linearly averages the
watermarked model with another model. Despite this, SEAL still successfully
extracts the embedded watermark bits, achieving a maximum BER of only 4.10\%
when extracting 256-bit watermarks in the black-box setting. These results
demonstrate SEAL's strong robustness against all evaluated attacks while
maintaining accurate and reliable multi-bit watermark extraction.

\noindent\textbf{SEAL Maintains Robustness in Both White-Box and Black-Box
Scenarios.} In Table~\ref{tab:rq2}, SEAL achieves nearly perfect extraction
accuracy across all evaluated attack types, with the highest BER being only
0.95\% in the white-box setting. In the black-box setting, attempts to remove
the inserted watermarks are slightly more effective, but still fail to
meaningfully affect the extraction accuracy. Under SFT attacks, SEAL attains an
average BER of 0.17\% in the white-box setting, while the average BER in the
black-box setting increases slightly to 1.21\%, which typically corresponds to
only one or two incorrectly extracted bits out of a total of 128 or 256 bits.
This minor increase in BER is expected, as the black-box setting provides only
output logits, which are more sensitive to changes in model parameters.
Nevertheless, SEAL demonstrates strong robustness against all evaluated attack
types under both white-box and black-box conditions.

\begin{figure}[h]
    \centering
    \includegraphics[width=0.95\linewidth]{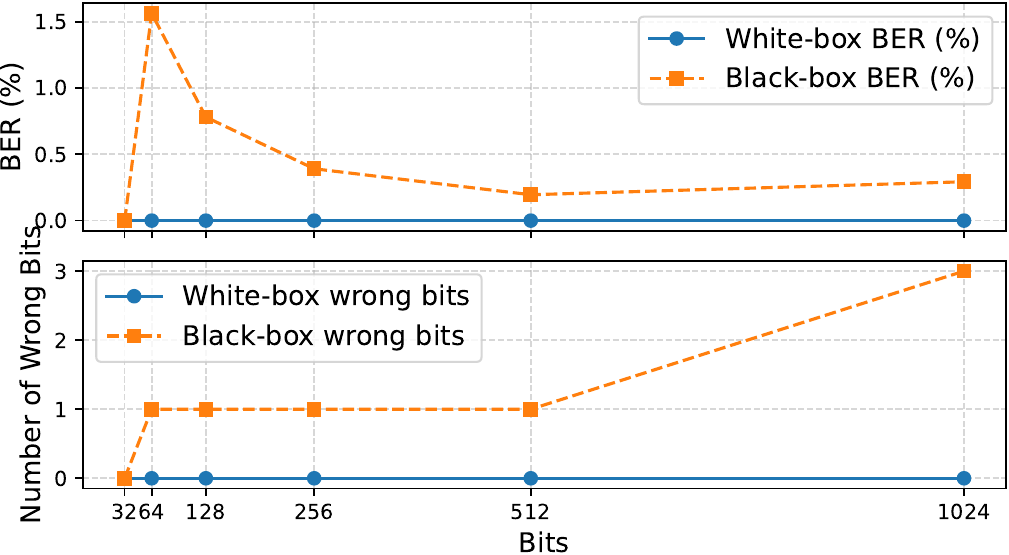}
    \caption{Scalability of SEAL in extracting multi-bit watermarks with
    increasing bit length under SFT attacks. We report the BER and the number of
    incorrectly extracted bits for each bit length.}
    \label{fig:rq2_capacity}
\end{figure}

\noindent\textbf{SEAL Supports Reliable Multi-Bit Watermark Extraction up to
1024 Bits.} We further examine the scalability of SEAL in extracting multi-bit
watermarks with increasing bit length under SFT attacks. As shown in
Fig.~\ref{fig:rq2_capacity}, SEAL can reliably embed and extract up to 1024
watermark bits. Since SEAL achieves perfect extraction performance in the
white-box setting, we focus our analysis on the black-box
setting to evaluate scalability.

As observed, the number of incorrectly extracted bits remains largely stable as
the total number of embedded bits increases. Specifically, no errors occur for
32-bit watermarks, while only one bit is extracted incorrectly when the total
bit length increases to 64, 256, or 512. When extracting 1024 bits, the number
of incorrectly extracted bits increases slightly to three. This results in an
interesting trend in BER. It first increases to 1.56\% at 64 bits and then
decreases to 0.78\% at 512 bits, as the number of incorrect bits remains
nearly constant while the total embedded bits grow exponentially.

These results demonstrate SEAL's strong scalability and robustness under
increased embedding capacity. While SEAL can theoretically support longer bit
sequences, we limit our experiments to 1024 bits due to practical considerations
in real-world watermark verification, where shorter bit lengths are typically
sufficient for copyright attribution.

\subsection{RQ3: Fidelity of Injecting SEAL}
We further evaluate the \textbf{Fidelity} of SEAL to examine whether watermark
injection degrades the model's original performance. Specifically, we compare
the unmodified model with its watermarked counterparts embedded with varying
numbers of verification bits across 15 downstream tasks, which cover a broad
range of language understanding capabilities, including:


\begin{itemize}[leftmargin=10pt]
  \item \textbf{General Knowledge and Reasoning:}
  MMLU~\cite{hendrycks2021measuringmassivemultitasklanguage},
  BoolQ~\cite{clark2019boolq}, Winogrande and WSC~\cite{sakaguchi2021winogrande},
  COPA~\cite{roemmele2011choice}, PiQA~\cite{bisk2020piqa},
  OpenBookQA~\cite{mihaylov2018can} and ARC Challenge~\cite{clark2018think}.

  \item \textbf{Sentence-level Understanding and Inference:}
  SST-2~\cite{wang2018glue}, CoLA~\cite{warstadt-etal-2019-neural},
  MRPC~\cite{dolan-brockett-2005-automatically}, RTE and
  NLI~\cite{wang2018glue}.

  \item \textbf{Lexical and Contextual Semantics:}
  CB~\cite{de2019commitmentbank}, WiC~\cite{pilehvar2018wic}.
\end{itemize}

We report the mean and standard deviation of performance for three
representative LLMs: LLaMA2-7B, LLaMA3-8B, and DeepSeek-Chat-7B. To examine the
effect of watermark injection at different bit lengths on the models' original
performance, we compare the baseline models with their SEAL-watermarked
counterparts, with the number of injected bits ranging from 64 to 512.

\begin{table}[htbp]
  \centering
  \caption{Performance difference of injecting SEAL watermarks with different
  numbers of injected bits.} 
  \resizebox{0.95\columnwidth}{!}{
  \begin{tabular}{lcccc}
      \toprule
      \textbf{Metric} & \textbf{64 Bits} & \textbf{128 Bits} & \textbf{256 Bits} & \textbf{512 Bits} \\
      \midrule
      ARC   & 0.008\scriptsize$\pm$0.002 & 0.013\scriptsize$\pm$0.003 & 0.029\scriptsize$\pm$0.002 & 0.055\scriptsize$\pm$0.014 \\
      BooLQ & 0.009\scriptsize$\pm$0.002 & 0.010\scriptsize$\pm$0.003 & 0.041\scriptsize$\pm$0.027 & 0.041\scriptsize$\pm$0.008 \\
      COPA  & 0.010\scriptsize$\pm$0.000 & 0.017\scriptsize$\pm$0.005 & 0.017\scriptsize$\pm$0.005 & 0.017\scriptsize$\pm$0.009 \\
      MMLU & 0.045\scriptsize$\pm$0.023 & 0.064\scriptsize$\pm$0.026 & 0.049\scriptsize$\pm$0.017 & 0.097\scriptsize$\pm$0.058 \\
      O.B.QA  & 0.007\scriptsize$\pm$0.005 & 0.019\scriptsize$\pm$0.005 & 0.015\scriptsize$\pm$0.010 & 0.036\scriptsize$\pm$0.030 \\
      PiQA  & 0.005\scriptsize$\pm$0.003 & 0.008\scriptsize$\pm$0.002 & 0.013\scriptsize$\pm$0.004 & 0.025\scriptsize$\pm$0.011 \\
      Wino.  & 0.004\scriptsize$\pm$0.003 & 0.004\scriptsize$\pm$0.004 & 0.013\scriptsize$\pm$0.003 & 0.016\scriptsize$\pm$0.007 \\
      WSC   & 0.032\scriptsize$\pm$0.027 & 0.022\scriptsize$\pm$0.032 & 0.100\scriptsize$\pm$0.095 & 0.151\scriptsize$\pm$0.045 \\
      \midrule
      NLI   & 0.024\scriptsize$\pm$0.021 & 0.052\scriptsize$\pm$0.035 & 0.069\scriptsize$\pm$0.052 & 0.066\scriptsize$\pm$0.043 \\
      RTE   & 0.014\scriptsize$\pm$0.005 & 0.042\scriptsize$\pm$0.017 & 0.006\scriptsize$\pm$0.003 & 0.048\scriptsize$\pm$0.018 \\
      MRPC  & 0.057\scriptsize$\pm$0.022 & 0.091\scriptsize$\pm$0.002 & 0.068\scriptsize$\pm$0.018 & 0.064\scriptsize$\pm$0.040 \\
      CoLA  & 0.045\scriptsize$\pm$0.011 & 0.040\scriptsize$\pm$0.023 & 0.038\scriptsize$\pm$0.024 & 0.124\scriptsize$\pm$0.016 \\
      SST   & 0.031\scriptsize$\pm$0.023 & 0.079\scriptsize$\pm$0.097 & 0.062\scriptsize$\pm$0.066 & 0.039\scriptsize$\pm$0.025 \\
      \midrule
      CB    & 0.030\scriptsize$\pm$0.008 & 0.012\scriptsize$\pm$0.008 & 0.060\scriptsize$\pm$0.015 & 0.074\scriptsize$\pm$0.037 \\
      WiC   & 0.009\scriptsize$\pm$0.008 & 0.012\scriptsize$\pm$0.009 & 0.013\scriptsize$\pm$0.009 & 0.011\scriptsize$\pm$0.009 \\
      \midrule
      \textbf{Avg} & \textbf{0.022\scriptsize$\pm$0.016} & \textbf{0.032\scriptsize$\pm$0.027} & \textbf{0.039\scriptsize$\pm$0.027} & \textbf{0.058\scriptsize$\pm$0.039} \\
      \bottomrule
  \end{tabular}
  }
  \label{tab:harmlessness}
\end{table}

\noindent\textbf{Injecting SEAL Preserves Model Performance.}
Table~\ref{tab:harmlessness} summarizes the performance differences between the
watermarked models and their original counterparts. As shown, injecting SEAL
watermarks has only a negligible impact on downstream performance across all
tasks and models, with the maximum average deviation being 0.058 and nearly all
task-wise differences below 0.1. In particular, when injecting 64-bit
watermarks, the performance of the watermarked models on general knowledge and
reasoning benchmarks remains almost identical to that of the original models.
Among the eight evaluated tasks, only MMLU and WSC exhibit slightly larger
differences (0.045 and 0.032, respectively), while all others show differences
below 0.01. These results highlight the lightweight nature of the watermarking
process based on model editing, which preserves the models' original
capabilities with only minimal performance variation.

\noindent\textbf{Aligning Anchor Samples with Bit Vectors Maintains
Sentence-Level and Lexical Understanding Utility.} As the process involves
aligning the hidden representations of anchor samples with corresponding bit
vectors, we examine whether injecting SEAL watermarks affects the models'
sentence-level and lexical understanding capabilities. As shown in
Table~\ref{tab:harmlessness} (Line 9-15), aligning anchor embeddings with
specific bit vectors has only a minimal impact on both sentence-level and
lexical understanding tasks. When injecting 64-bit watermarks, the performance
differences across all evaluated benchmarks remain within 0.06, with the largest
deviation observed on the MRPC task (0.057). These results further confirm that
the alignment process is effectively harmless to the models' original
understanding capabilities.

\noindent\textbf{SEAL Maintains Model Utility Across Different Watermark
Lengths.} We further investigate whether increasing the number of injected
watermark bits affects the models' original performance. As shown in
Table~\ref{tab:harmlessness}, although the average performance difference
slightly increases with the watermark length, it causes negligible influence to
the original models' capabilities. Specifically, the average deviation rises
modestly from 0.022 to 0.058 as the number of injected bits increases from 64 to
512. This behavior is expected, since embedding longer bit sequences requires
aligning more anchor samples with bit vectors, which may introduce minimal noise
into the model representations. Nevertheless, even at 512 bits, the observed
degradation is marginal, further confirming that SEAL effectively preserves
model utility across different watermark capacities.

\subsection{RQ4: SEAL Efficiency}
We further evaluate the \textbf{Efficiency} of SEAL in terms of both runtime and
memory usage, and compare it against existing watermarking and fingerprinting
methods. For fair comparison, we first measure efficiency by inserting and
verifying a 1-bit watermark, as prior methods are limited to carrying a single
bit of information. We then analyze runtime scalability with respect to the
number of embedded bits.

\begin{table}[htbp]
    \centering
    \caption{Runtime comparison of watermark inserting and verification. ``All''
    reports the total runtime (in seconds).} \resizebox{0.9\columnwidth}{!}{
      \begin{tabular}{ccccc}
      \toprule
      \textbf{Evaluation} & \textbf{Methods} & \textbf{Insertion} & \textbf{Verification} & \textbf{All} \\
      \midrule
      \multirow{4}[2]{*}{White-box} & Gradient & - & 28.25 & 28.25 \\
            & Huref & -     & 62.94 & 62.94 \\
            & Reef  & -     & 13.94 & 13.94 \\
            & SEAL(W) & 34.32 & 0.027 & 34.35 \\
      \midrule
      \multirow{5}[2]{*}{Black-box} & Trap  & -     & 3391.18 & 3391.18 \\
            & MET   & -     & 48.01 & 48.01 \\
            & LLMMap & -     & 39.72 & 39.72 \\
            & IF    & 20.72 & 18.80  & 39.52 \\
            & SEAL(B) & 34.82 & 1.129 & 35.93 \\
      \bottomrule
      \end{tabular}%
    }
    \label{tab:rq4_runtime}%
  \end{table}%

\noindent\textbf{SEAL is Efficient to Embed and Verify in Runtime.} As shown in
Table~\ref{tab:rq4_runtime}, SEAL provides high runtime efficiency in both
white-box and black-box settings. In the white-box case, SEAL completes
insertion and verification in 34.35 seconds, which is comparable to Gradient and
nearly $2\times$ faster than Huref. Although Reef verifies faster (13.94
seconds), it does not perform watermark insertion. Once embedded, SEAL verifies
in only 0.027 seconds, nearly 500$\times$ faster than Huref. In the black-box
setting, SEAL also outperforms existing methods. It completes insertion and
verification in 35.93 seconds, about 4 seconds faster than IF and LLMMap and 13
seconds faster than MET, while Trap requires over 3000 seconds. Overall, SEAL
offers a balanced, scalable, and low-latency runtime design across both access
regimes.

\begin{figure}[h]
    \centering
    \includegraphics[width=0.95\linewidth]{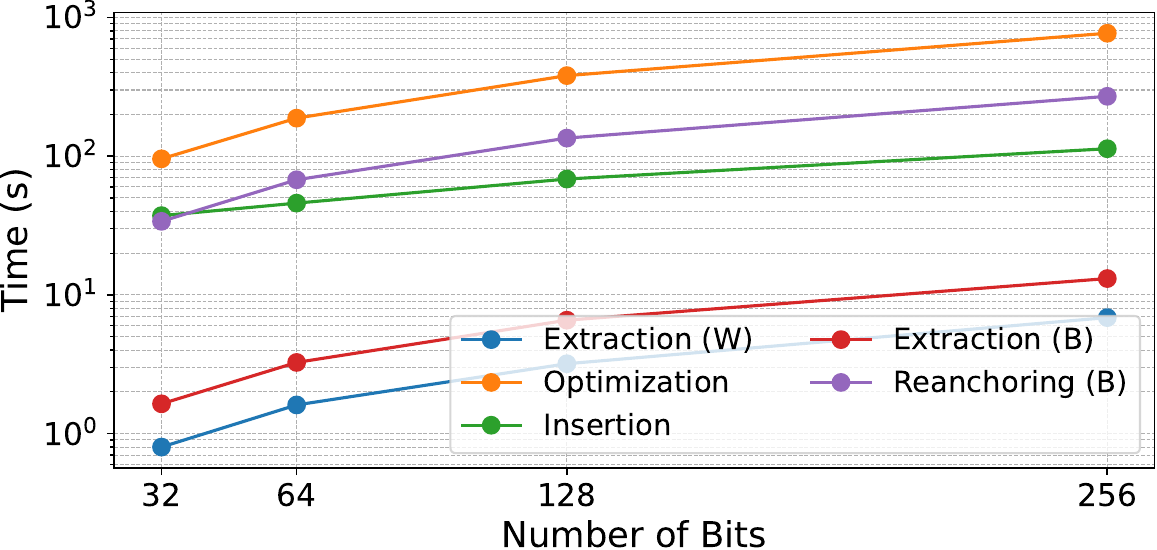}
    \caption{Efficiency study of watermark insertion and verification for
    different bit lengths in runtime.}
    \label{fig:seal_times}
\end{figure}

We analyze SEAL's runtime scalability with respect to the number of embedded
bits. The pipeline is divided into four stages: (1) optimization of anchor-bit
alignment, (2) insertion of parameter updates, (3) reanchoring of sentinel
weights, and (4) extraction of watermark bits. Figure~\ref{fig:seal_times}
reports the runtime of each stage under both white-box and black-box settings.

The insertion stage behaves similarly across settings, while extraction differs:
black-box extraction operates on output logits, whereas white-box extraction
directly reads latent representations. Reanchoring is required only in the
black-box setting and adds modest overhead.

\noindent\textbf{SEAL is Scalable to Embed Multiple Bits.}
Figure~\ref{fig:seal_times} shows how SEAL's runtime scales with watermark size.
The anchor-bit optimization stage dominates the cost, accounting for over half
of the total runtime (e.g., 188.11 seconds for a 64-bit watermark, compared to
116.77 seconds for all other stages combined). This cost can be further reduced
by performing optimization \textit{offline} on batches of anchor candidates.

The insertion stage scales near-sublinearly with the number of bits, and
extraction is inexpensive (6.86 seconds and 13.12 seconds for a 256-bit
watermark in white-box and black-box settings). The reanchoring stage, required
only for black-box verification, adds overhead but remains far cheaper than
optimization. Overall, SEAL provides efficient and scalable multi-bit embedding
and verification.

\begin{figure}[h]
    \centering
    \includegraphics[width=\linewidth]{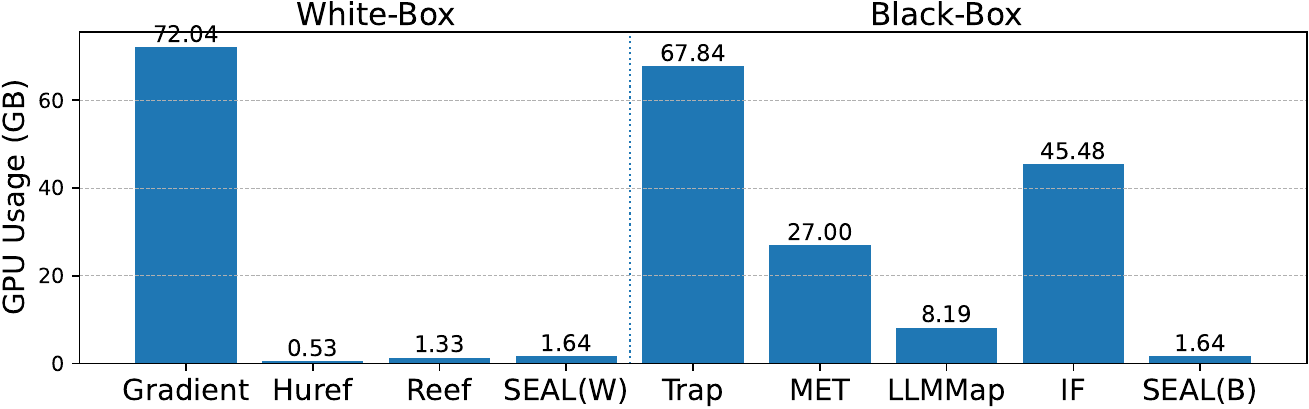}
    \caption{GPU memory usage (GB) during watermark insertion and verification.}
    \label{fig:seal_gpu}
\end{figure}

\noindent\textbf{SEAL Watermark is GPU-Efficient.} As shown in
Figure~\ref{fig:seal_gpu}, SEAL requires only about 1.6 GB of GPU memory in both
white- and black-box settings, which is comparable to Reef and Huref and far
lower than Gradient (72 GB). In the black-box case, SEAL also remains at 1.6 GB,
whereas methods such as Trap and IF consume up to 68 GB and 45 GB. These results
show that SEAL offers fast embedding and verification with an extremely small
GPU footprint, making it practical for real-world deployment.

\subsection{RQ5: Hyperparameter Analysis}
\label{sec:hyperparameters}
\begin{table}[htbp]
  \centering
  \caption{Effectiveness of the reanchoring mechanism.}
  \resizebox{0.95\columnwidth}{!}{
    \begin{tabular}{ccccccc}
    \toprule
    \multirow{2}[2]{*}{\textbf{Models}} 
      & \multicolumn{3}{c}{\textbf{Without Reanchoring}} 
      & \multicolumn{3}{c}{\textbf{With Reanchoring}} \\
    \cmidrule(lr){2-4} \cmidrule(lr){5-7}
      & \textbf{BER}$\downarrow$   & \textbf{BSE}$\uparrow$ & \textbf{BST}$\downarrow$ & \textbf{BER}$\downarrow$   & \textbf{BSE}$\uparrow$ & \textbf{BST}$\downarrow$ \\
    \midrule
    LLaMA2-7B     & 6.250 & 0.004 & 0.003 & 1.563 & 0.005 & 0.002 \\
    LLaMA3-8B     & 15.625 & 0.007 & 0.004 & 1.563 & 0.016 & 0.007 \\
    Qwen2.5-7B    & 7.813 & 0.004 & 0.002 & 0.000 & 0.006 & 0.001 \\
    Mistral-7B    & 4.688 & 0.008 & 0.005 & 1.563 & 0.010 & 0.004 \\
    Deepseek-7B   & 10.938 & 0.005 & 0.001 & 0.000 & 0.007 & 0.001 \\
    \bottomrule
    \end{tabular}%
  }
  \label{tab:rq5_reanchor}%
\end{table}%

\noindent\textbf{Effectiveness of Black-Box Reanchoring.}
Table~\ref{tab:rq5_reanchor} shows that reanchoring is crucial for accurate
black-box extraction. Without it, BER ranges from 4.688\% to 15.625\% across
LLMs. After reanchoring the sentinel weights, BER drops to 1.563\% (one error in
64 bits) or even 0.000\%. Reanchoring also improves bit separability. For
LLaMA3-8B, BSE increases from 0.007 to 0.016 (more than 2$\times$). These
results demonstrate that reanchoring restores the impact of bit vectors on
output logits even after fine-tuning, enabling reliable black-box watermark
recovery.

\begin{table}[htbp]
  \centering
  \caption{Impact of anchor-sample selection.}
  \setlength{\tabcolsep}{2.5pt}
  \resizebox{0.9\columnwidth}{!}{
    \begin{tabular}{cccccccc}
    \toprule
    \multirow{2}[2]{*}{\textbf{Evaluation}} & \multirow{2}[2]{*}{\textbf{Selection}} &
    \multicolumn{3}{c}{\textbf{Before SFT}} & \multicolumn{3}{c}{\textbf{After SFT}} \\
    \cmidrule(lr){3-5}\cmidrule(lr){6-8}
          &       & \textbf{BER}$\downarrow$   & \textbf{BSE}$\uparrow$ & \textbf{BST}$\downarrow$ & \textbf{BER}$\downarrow$   & \textbf{BSE}$\uparrow$ & \textbf{BST}$\downarrow$ \\
    \midrule
    \multirow{3}[2]{*}{White-box} & High-confidence  & 0.000     & 0.058 & 0.782 & 0.000     & 0.041 & 0.376 \\
          & Low-confidence  & 0.000     & 0.036 & 0.620 & 0.000     & 0.024 & 0.263 \\
          & Random & 0.000     & 0.036 & 0.665 & 0.000     & 0.027 & 0.273 \\
    \midrule
    \multirow{3}[2]{*}{Black-box} & High-confidence  & 0.000     & 0.012 & 0.002 & 1.563 & 0.005 & 0.002 \\
          & Low-confidence  & 0.000     & 0.009 & 0.002 & 15.625 & 0.003 & 0.002 \\
          & Random & 0.000     & 0.010 & 0.002 & 6.250  & 0.003 & 0.002 \\
    \bottomrule
    \end{tabular}%
  }
  \label{tab:rq5_anchor}%
\end{table}

\noindent\textbf{Selecting High-Confidence Anchors Improves Watermark
Robustness.} To justify our choice of anchor samples for watermark embedding, we
evaluate three strategies for selecting $(\textit{subject}, \textit{relation},
\textit{object})$ triplets from the CounterFact dataset during SEAL injection.
The \textit{high-confidence} strategy selects samples whose objects already have
the highest model-predicted probabilities; the \textit{low-confidence} strategy
selects those with the lowest probabilities. \textit{Random} sampling serves as
a reference. 

As shown in Table~\ref{tab:rq5_anchor}, using high-confidence samples yields the
most reliable extraction under SFT in both white- and black-box settings. In the
black-box case, this strategy attains a post-SFT BER of 1.563\%, whereas
low-confidence and random samples lead to much higher BERs (15.625\% and
6.250\%). This supports using high-confidence samples as anchors, since
low-confidence ones bias the optimization toward fitting target outputs rather
than aligning latent embeddings with the bit vectors.

\begin{table}[htbp]
  \centering
  \caption{Watermark extraction across insertion layers.}
  \setlength{\tabcolsep}{2.5pt}
  \resizebox{0.9\columnwidth}{!}{
    \begin{tabular}{cccccccc}
    \toprule
    \multirow{2}[2]{*}{\textbf{Evaluation}} & \multirow{2}[2]{*}{\textbf{Insert Layers}} & \multicolumn{3}{c}{\textbf{Before SFT}} & \multicolumn{3}{c}{\textbf{After SFT}} \\
    \cmidrule(lr){3-5}\cmidrule(lr){6-8}
          &       & \textbf{BER}$\downarrow$   & \textbf{BSE}$\uparrow$ & \textbf{BST}$\downarrow$ & \textbf{BER}$\downarrow$   & \textbf{BSE}$\uparrow$ & \textbf{BST}$\downarrow$ \\
    \midrule
    \multirow{7}[2]{*}{white-box} & 30       & 0.000 & 0.228 & 0.750 & 0.000 & 0.181 & 0.590 \\
                                  & 15       & 0.000 & 0.035 & 0.651 & 0.000 & 0.061 & 0.605 \\
                                  & 5        & 0.000 & 0.005 & 0.639 & 0.000 & 0.003 & 0.564 \\
                                  & 28,29,30 & 0.000 & 0.008 & 0.772 & 1.563 & 0.003 & 0.484 \\
                                  & 13,14,15 & 0.000 & 0.011 & 0.694 & 0.000 & 0.013 & 0.572 \\
                                  & 3,4,5    & 0.000 & 0.007 & 0.206 & 0.000 & 0.007 & 0.193 \\
                                  & 5,15,30  & 0.000 & 0.038 & 0.782 & 0.000 & 0.021 & 0.633 \\
    \midrule
    \multirow{7}[2]{*}{black-box} & 30       & 0.000 & 0.013 & 0.002 & 7.813 & 0.006 & 0.003 \\
                                  & 15       & 14.063 & 0.005 & 0.003 & 32.813 & 0.004 & 0.003 \\
                                  & 5        & 35.938 & 0.003 & 0.002 & 48.438 & 0.002 & 0.002 \\
                                  & 28,29,30 & 0.000 & 0.011 & 0.003 & 4.688 & 0.004 & 0.003 \\
                                  & 13,14,15 & 23.438 & 0.004 & 0.003 & 28.125 & 0.003 & 0.002 \\
                                  & 3,4,5    & 40.625 & 0.001 & 0.001 & 45.313 & 0.002 & 0.001 \\
                                  & 5,15,30  & 0.000 & 0.012 & 0.002 & 1.563 & 0.005 & 0.002 \\
    \bottomrule
    \end{tabular}%
  }
  \label{tab:rq5_layers}%
\end{table}%

\noindent\textbf{Selecting Proper Insertion Layers Enhances Watermark
Robustness.} We examine how different configurations of insertion layers affect
watermark extraction. We compare single-layer and multi-layer strategies by
embedding SEAL into transformer layers at various depths, and evaluate watermark
extraction performance before and after SFT.

As shown in Table~\ref{tab:rq5_layers}, all configurations achieve perfect
white-box extraction (BER $=0$) before SFT, with only minimal degradation
afterward. In contrast, layer choice matters substantially in the black-box
setting. Watermarks inserted into lower layers are far more fragile, yielding
high BER even before SFT (e.g., 35.94\% at layer (5) and 40.63\% at (3,4,5)),
whereas deeper layers such as (15) or (30) reduce BER to 14.06\% and 0.00\%.
This is because black-box verification depends on the watermark's influence on
output logits, and lower-layer modifications have limited effect on the final
output space.

We also find that distributing watermark updates across multiple layers improves
robustness to SFT. For instance, spreading updates across (28,29,30) or
(13,14,15) yields post-SFT BERs of 4.69\% and 28.13\%, compared to 7.81\% and
14.06\% with single-layer insertion. Combining layers at different depths
provides the strongest resilience: inserting into (5,15,30) reduces post-SFT BER
to 0.00\%. This demonstrates that multi-layer, cross-depth embedding
significantly enhances SEAL's robustness under fine-tuning and black-box
verification.

\section{Robustness to Knowledgeable Attackers}
\label{sec:adaptive_attacks}
We evaluate a strong, knowledgeable adversary who (i) knows the SEAL mechanism and
the set of layers used for insertion and (ii) attempts to erase the watermark by
fine-tuning only those layers. 

\begin{table}[htbp]
  \centering
  \caption{Impact of insertion layer selection.}
  \resizebox{0.9\columnwidth}{!}{
    \begin{tabular}{cccccccc}
    \toprule
    \multirow{2}[2]{*}{\textbf{Evaluation}} & \multirow{2}[2]{*}{\textbf{Attack Layers}} &
    \multicolumn{3}{c}{\textbf{Before SFT}} & \multicolumn{3}{c}{\textbf{After SFT}} \\
    \cmidrule(lr){3-5}\cmidrule(lr){6-8}
          &       & \textbf{BER}$\downarrow$   & \textbf{BSE}$\uparrow$ & \textbf{BST}$\downarrow$ & \textbf{BER}$\downarrow$   & \textbf{BSE}$\uparrow$ & \textbf{BST}$\downarrow$ \\
    \midrule
    \multirow{4}[2]{*}{white-box} & 26,27,28,29,30 & 0.000 & 0.158 & 0.775 & 0.000 & 0.133 & 0.625 \\
          & 11,12,13,14,15 & 0.000 & 0.037 & 0.788 & 0.000 & 0.011 & 0.449 \\
          & 10,9,8,7,6,5 & 0.000 & 0.010 & 0.787 & 0.000 & 0.002 & 0.250 \\
          & 5,15,30 & 0.000 & 0.187 & 0.763 & 0.000 & 0.156 & 0.636 \\
    \midrule
    \multirow{4}[2]{*}{black-box} & 26,27,28,29,30 & 0.000 & 0.011 & 0.002 & 0.000 & 0.006 & 0.002 \\
          & 11,12,13,14,15 & 0.000 & 0.011 & 0.002 & 3.125 & 0.004 & 0.002 \\
          & 10,9,8,7,6,5 & 0.000 & 0.011 & 0.002 & 1.563 & 0.005 & 0.002 \\
          & 5,15,30 & 0.000 & 0.012 & 0.002 & 0.000 & 0.006 & 0.002 \\
    \bottomrule
    \end{tabular}%
  \label{tab:rq_adaptive}%
  }
\end{table}%

\noindent\textbf{Knowledgeable fine-tuning on known insertion layers does not remove
SEAL.} Table~\ref{tab:rq_adaptive} shows that BER remains zero in all white-box
cases, indicating that targeted updates cannot eliminate the multi-bit signal
when hidden states are available for verification. In the black-box setting, BER
increases only slightly for a few layer groups (e.g., up to 3.125), while
high-level or cross-depth insertions (e.g., \([30,15,5]\)) remain fully
recoverable (BER $=0$).

This robustness stems from two factors. First, SEAL distributes watermark
information across semantic subspaces and multiple depths, making localized
fine-tuning insufficient to remove the signal. Second, adversaries lack the
defender's secret anchor samples and optimization targets, so their fine-tuning
cannot replicate the alignment required to erase the watermark without harming
model utility. Thus, even knowing the insertion layers is insufficient: without
access to the hidden anchors, no practical fine-tuning attack can reliably
eliminate SEAL's multi-bit watermark.

%% file: docs/07-conclusion.tex
\section{Conclusion}
We presented SEAL, a subspace-anchored watermarking framework that embeds
multi-bit signatures into the latent space of LLMs without altering their
behavior. By aligning factual anchor representations with orthogonal bit
vectors, SEAL achieves functionally harmless and persistent watermarking
resilient to various model modifications. It supports both white-box
verification via hidden-state inspection and black-box verification via Bayesian
reanchoring, which reliably recovers watermarks under model drift. Experiments
show that SEAL achieves near-perfect lineage identification (AUC = 1.00) and
very low bit-error rates ($< 1.21$ \%) while maintaining model utility, making
it a practical solution for LLM ownership protection.

%% file: docs/08-appendices.tex
\appendices

\input{docs/081-compare_pool.tex}

\input{docs/082-reanchoring_proof.tex}

\input{docs/083-ecc.tex}

\input{docs/084-metric_detail.tex}

%% file: docs/081-compare_pool.tex
\section{Comparison Pool}
We list the comparison pool used for model lineage identification in Table~\ref{tab:comparepool}.
\begin{table*}[t]
    \centering
    \small
    \caption{Comparison pool used for model lineage identification. Each column
    represents a target model, and each row lists the candidate models evaluated for
    determining whether they are derivatives of the corresponding target model.} 
    \resizebox{\textwidth}{!}{
    \begin{tabular}{l|l|l|l|l|l}
    \toprule
    \textbf{LLaMA2-7B} & \textbf{Qwen2.5-7B} & \textbf{LLaMA3-8B} & \textbf{Mistral-7B} & \textbf{Deepseek-7B} & \textbf{Qwen2.5-14B} \\
    \midrule
    LLaMA2-7B (PEFT) - alpaca & Qwen2.5-7B (PEFT) - alpaca & LLaMA3-8B (PEFT) - alpaca & Mistral-7B (PEFT) - alpaca & Deepseek-7B (PEFT) - alpaca & qwen-14b (PEFT) - alpaca \\
    LLaMA2-7B (PEFT) - shareGPT & Qwen2.5-7B (PEFT) - shareGPT & LLaMA3-8B (PEFT) - shareGPT & Mistral-7B (PEFT) - shareGPT & Deepseek-7B (PEFT) - shareGPT & qwen-14b (PEFT) - shareGPT \\
    LLaMA2-7B (SFT) - alpaca & Qwen2.5-7B (SFT) - alpaca & LLaMA3-8B (SFT) - alpaca & Mistral-7B (SFT) - alpaca & Deepseek-7B (SFT) - alpaca & qwen-14b (SFT) - alpaca \\
    LLaMA2-7B (SFT) - shareGPT & Qwen2.5-7B (SFT) - shareGPT & LLaMA3-8B (SFT) - shareGPT & Mistral-7B (SFT) - shareGPT & Deepseek-7B (SFT) - shareGPT & qwen-14b (SFT) - shareGPT \\
    LLaMA2-7B (DT) - LLaMA & Qwen2.5-7B (DT) - alpaca & LLaMA3-8B (DT) - alpaca & Mistral-7B (DT) - alpaca & Deepseek-7B (DT) - alpaca & qwen-14b (DT) - alpaca \\
    LLaMA2-7B (DT) - shareGPT & Qwen2.5-7B (DT) - shareGPT & LLaMA3-8B (DT) - shareGPT & Mistral-7B (DT) - shareGPT & Deepseek-7B (DT) - shareGPT & qwen-14b (DT) - shareGPT \\
    Qwen2.5-7B & LLaMA2-7B & Qwen2.5-7B & Qwen2.5-7B & Qwen2.5-7B & Qwen2.5-7B \\
    LLaMA3-8B & LLaMA3-8B & LLaMA2-7B & LLaMA3-8B & LLaMA3-8B & LLaMA3-8B \\
    LLaMA2-13B & LLaMA2-13B & LLaMA2-13B & LLaMA2-13B & LLaMA2-13B & LLaMA2-13B \\
    Mistral-7B & Mistral-7B & Mistral-7B & LLaMA2-7B & Mistral-7B & Mistral-7B \\
    Deepseek-7B & Deepseek-7B & Deepseek-7B & Deepseek-7B & LLaMA2-7B & Deepseek-7B \\
    Gemma-7B & Gemma-7B & Gemma-7B & Gemma-7B & Gemma-7B & Gemma-7B \\
    \bottomrule
    \end{tabular}
    }
    \label{tab:comparepool}
    \end{table*}

%% file: docs/082-reanchoring_proof.tex
\section{Derivation of Bayesian Reanchoring}
\label{sec:bayes_reanchor_derivation}
We derive the closed-form posterior mean used in our Bayesian reanchoring
procedure in \Cref{eq:bayes_reanchor_reorder}, under standard Gaussian
assumptions.

\noindent\textbf{Problem Setup.}
Let $w \in \mathbb{R}^m$ denote the unknown sentinel projection vector under the
current (potentially fine-tuned) model.  
We assume access to:
\begin{itemize}[leftmargin=1em]
    \item A prior estimate $w_{\text{sig}}$ computed during the white-box stage.
    \item A sample mean drift $\mu_\Delta$ estimated from paired logit
    differences $\Delta \mathcal{L}_k = L_k^{\text{D}} - L_k^{\text{W}}$ across
    $K$ reference prompts.
\end{itemize}

Our goal is to infer the most likely value of $w$ given the observed drift
$\mu_\Delta$.

\noindent\textbf{Bayesian Formulation.}
We place the following Gaussian prior on $w$:
\begin{equation}
p(w) = \mathcal{N}(w_{\text{sig}}, \lambda^{-1} I),
\end{equation}
and model the likelihood of the observed logit shift as:
\begin{equation}
p(\mu_\Delta \mid w) = \mathcal{N}(w - w_{\text{sig}}, \rho \Sigma_\Delta),
\end{equation}
which reflects the assumption that the empirical drift $\mu_\Delta$ is a noisy
observation of the true shift $w - w_{\text{sig}}$.

Applying Bayes' rule, the posterior is proportional to:
\begin{equation}
p(w \mid \mu_\Delta) \propto p(\mu_\Delta \mid w) \cdot p(w).
\end{equation}

Taking the negative log-posterior (up to constants), we obtain the following
objective:
\begin{align}
\mathcal{L}(w) &= \frac{1}{2\rho} ( \mu_\Delta - (w - w_{\text{sig}}))^\top \Sigma_\Delta^{-1} ( \mu_\Delta - (w - w_{\text{sig}})) \nonumber \\
&+\frac{\lambda}{2} \|w - w_{\text{sig}}\|_2^2.
\end{align}

\noindent\textbf{Closed-Form Solution (MAP).}
Expanding the quadratic terms and minimizing $\mathcal{L}(w)$ yields the maximum
a posteriori (MAP) estimate:
\begin{equation}
(\rho \Sigma_\Delta + \lambda I) w = \rho \Sigma_\Delta (\mu_\Delta + w_{\text{sig}}) + \lambda w_{\text{sig}}.
\end{equation}

Solving for $w$ gives:
\begin{equation}
w = (\rho \Sigma_\Delta + \lambda I)^{-1} \left[ \rho \Sigma_\Delta (\mu_\Delta + w_{\text{sig}}) + \lambda w_{\text{sig}} \right].
\end{equation}

To remove uniform offsets (which do not affect verification), we apply a
zero-mean projection:
\begin{equation}
P = I - \frac{\mathbf{1}\mathbf{1}^\top}{m},
\end{equation}
and express the final reanchored vector as:
\begin{equation}
w^* = (\rho \Sigma_\Delta + \lambda I)^{-1} (w_{\text{sig}} + P \mu_\Delta + \lambda P w_{\text{sig}}),
\end{equation}
which is the form used in \Cref{eq:bayes_reanchor_reorder}.

%% file: docs/083-ecc.tex
\section{Error Correction Coding (ECC) for Enhanced Reliability}
\label{sec:ecc}
Although SEAL already achieves high watermark extraction accuracy
under both white-box and black-box settings, further reliability can be achieved
by integrating ECC into the bit-level representation. Specifically, each $k$-bit
watermark segment can be encoded into an $n$-bit ECC codeword (e.g., using
Hamming~\cite{hamming1950error} or BCH codes~\cite{bose1960class}), which allows
the verifier to correct a small number of bit flips during extraction. We
anticipate that, this design is particularly beneficial under noisy or partially
compromised conditions, where quantization, model merging, or low-bit retrieval
may induce minor bit errors.

Notice that integrating ECC does not change SEAL's embedding procedure or
security assumptions. The encoder simply maps the original watermark bits into a
redundant but error-tolerant format before optimization, and the verifier
applies standard ECC decoding upon recovery. In practice, a lightweight $(n, k)$
configuration such as $(128, 64)$ provides strong correction capability with
less than 2$\times$ bit overhead, while maintaining the same optimization cost
and watermark invisibility. This enhancement ensures that SEAL remains
verifiable even when a small portion of the embedded bits are perturbed, further
improving robustness against extreme post-deployment model transformations.

%% file: docs/084-metric_detail.tex
\section{Details of Evaluation Metrics for Lineage Identification}
\label{sec:metric_detail_li}
We introduce the details of the evaluation metrics used in evaluating the
effectiveness of model lineage identification.
\begin{itemize}[leftmargin=1em]
    \item \textbf{Area Under the ROC Curve (AUC).} This metric quantifies the
    overall capability of distinguishing between \textit{derivative} and
    \textit{independent} models. A higher AUC indicates stronger discriminative
    power.
    
    \item \textbf{Partial AUC (pAUC).} This metric measures the performance
    under a low false positive rate (FPR) regime, typically in the range $[0,
    0.05]$, focusing on the method's robustness against false positives.
    
    \item \textbf{Mahalanobis Distance (MD).} This metric computes the mean
    distance between the derived score distributions of derivative and
    independent models. A higher MD reflects better separation between model
    lineages.
\end{itemize}